\documentclass[conference]{IEEEtran}
\ifCLASSINFOpdf
  \usepackage[pdftex]{graphicx}
\else
\fi
\usepackage[utf8]{inputenc}
\usepackage{amsthm}
\usepackage{amsmath, amssymb}
\usepackage{url}
\usepackage{fancyvrb}
\usepackage{textcomp}
\usepackage{verbatim}
\usepackage{fancyvrb}
\usepackage{color}
\usepackage{listings}
\usepackage{mathpartir}
\usepackage{cleveref}
\crefname{subsection}{§}{§§}
\Crefname{section}{§}{§§}
\newtheorem{thm}{Theorem}
\newtheorem{lem}{Lemma}
\begin{document}
%\title{Proof of Soundness and Completeness of an security monitor}
\title{Sound and Complete Runtime Security Monitor for Application Software}

\author{\IEEEauthorblockN{Muhammad Taimoor Khan}
\IEEEauthorblockA{QCRI, HBKU}
\and
\IEEEauthorblockN{Dimitrios Serpanos}
\IEEEauthorblockA{QCRI, HBKU}
\and
\IEEEauthorblockN{Howard Shrobe}
\IEEEauthorblockA{MIT CSAIL, USA}}
\maketitle

\begin{abstract}
Conventional approaches for ensuring the security of application software at run-time, through monitoring, either produce (high rates 
of) false alarms (e.g. intrusion detection systems) or 
limit application performance (e.g. run-time verification). 
We present a run-time security monitor that detects both known and
unknown cyber attacks by checking that the run-time behavior of the
application is consistent with the expected behavior modeled in application specification. This is crucial because, even if the implementation is
consistent with its specification, the application may still be vulnerable
due to flaws in the supporting infrastructure (e.g. the language
run-time system, supporting libraries and the operating system). This run-time security monitor is sound and complete, eliminating false alarms, as well as efficient, so that it does not limit run-time application performance and so that it supports real-time systems. Importantly, this monitor is readily applicable to both legacy and new system platforms.

The security monitor takes as input the application specification and the application implementation, which may be expressed in different languages. 
The specification language of the application software is formalized based on monadic second order logic
(i.e. first order logic and set theory) and event calculus interpreted
over algebraic data structures. This language allows us to express behavior of an application at any desired (and practical) level of abstraction as well as with high degree of modularity. 
%Furthermore, it allows us to describe known or potential attack plans,
% facilitating early detection of more
%sophisticated attacks and thus, improving performance. 
%
The security monitor detects every
attack by systematically comparing the application execution and specification behaviors at runtime, even though they operate at two different levels
of abstraction. We define the denotational
semantics of the specification language and prove that the monitor is
sound and complete, i.e. if the application is consistent with its
specification, the security monitor will produce no false alarms 
(soundness) and that it will detect any deviation of the application
from the behavior sanctioned by the specification language (completeness). 
Furthermore, the monitor is efficient because of the modular
application
specification at appropriate level(s) of abstraction. Importantly,
the application specification language enables the description of known 
or potential attack plans, enabling not only attack detection but
attack characterization as well and, thus, facilitating effective 
and efficient defenses to sophisticated attacks.
% facilitating early detection of more
%sophisticated attacks and thus, improving performance. 

%The proof is essentially a
%structural induction on the elements of the specification language.  
%We illustrate this with an example of a simple Industrial Control System (ICS).
\end{abstract}


% \fi
\IEEEpeerreviewmaketitle

\section{Introduction}\label{sec:intro}
Runtime security monitors are components of defending systems against cyber attacks and must provide fast and accurate detection of attacks.  Conventional run-time monitoring systems suffer from high false alarm
rates, for both positive and negative alarms, and are inefficient because
their typical amount of observed parameters is large and possibly
irrelevant to a number of attacks. There are two 
key reasons for these limitations: first, the systems do not ``understand'' the complete behavior of the system they are protecting, and second, the systems do not ``understand'' what an attacker is 
trying to achieve.  Actually, most such systems are retrospective, taking into account and analyzing historical data, resulting to
attack surface signatures of previous attacks and attempting to
identify the same signature(s) in new traffic. Thus, conventional
run-time monitors are passive, waiting for (and expecting that)
something similar to what has already happened to recur. Attackers, 
of course, respond by varying their attacks so as to avoid detection.

\begin{figure}
\centering
\includegraphics[scale=0.55]{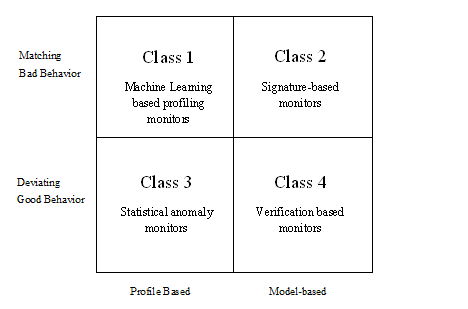}
\caption{Classification of Runtime Security Monitoring Systems}
\label{fig:quad}
\end{figure}
%Based on AWDRAT~\cite{Shrobe:2006}, ARMET is a representative of a new class of protection systems that employ a different, active form of perception, one that is informed both by knowledge of what the protected application is trying to do and by knowledge of how attackers think.  It employs both bottom-up reasoning (going from sensors data to conclusions about what attacks might be in progress) as well as top-down reasoning (given a set of hypotheses about what attacks might be in progress, it focuses its attention to those events most likely to significantly help in discerning the ground truth).

There are two dimensions along which run-time monitoring systems for
security can be classified. The first one is the behavior description
method, i.e. profile-based or model-based. The second one is the behavior comparison method, i.e. matching to bad behavior or deviation
from good behavior. This classification approach leads to four classes,
as shown in Figure~\ref{fig:quad}, which include existing techniques
and systems, each with its own strengths and weaknesses.
Profile-based systems that detect attacks by matching with bad behavior (Class 1
in the figure) typically employ statistical and machine learning 
methods to build a profile of bad behavior of the systems and more
specifically, build statistical profiles of attacks (e.g.,~\cite{Hodge:2004, Valdes:2000}). These systems are more robust than 
model based systems, since the machine learning techniques tend to
generalize from the data presented. However, they do not provide rich diagnostic information and suffer from false alarms.
Alternatively, profile-based systems that detect deviation from good
behavior (Class 3 in the figure) typically build a statistical profile
of normal (good) behavior and detect deviations from this profile (e.g.~\cite{Kim:2004, Lakhina:2005}). Such anomaly detectors are even more
robust than Class 1 systems, because they do not depend on historical
knowledge of the form of an attack. However, they have a significant
false alarm rate, because they have limited diagnostic information:
when a deviation is detected, the known information about it is that
something out of the ordinary has happened, but there is no sufficient
information whether this is malicious, accidental or just a variation
of the normal behavior beyond the statistically accepted profile.

Model-based systems (Classes 2 and 4 in Figure~\ref{fig:quad}) are
popular in highly secure environments, where successful attacks cause
significantly high costs. 
Signature-based systems are a typical 
example in this class~(e.g.~\cite{Vern:1998, Martin:1999}), and 
they look for matches to bad behavior, i.e. they are systems in Class 2.
The advantage of such systems is that, when
a match occurs, i.e. an attack is detected, the systems have enough
diagnostic information available to "understand" what the failure 
has been.
However, they lack robustness, since they will fail to detect an attack,
if they have no model of it; thus, they are susceptible to zero-day
attacks and, in general, attacks they have not been trained for. 
Finally, model-based systems that employ run-time software
verification to detect deviation from good behavior fall in Class 4 
of the figure. These systems model the good behavior of a system 
(e.g.~\cite{Watterson:2007, Zhang:2008}) and detect deviations from
that behavior using run-time software verification techniques. 
Their advantage is that, whenever the system execution deviates from
good behavior, there is knowledge of the exact problem that led to
the deviation (i.e. the offending instruction or routine). However,
such verification methods (a) require adequate design/implementation
information of the system to operate (which is usually not the case 
for legacy systems) and (b) limit run-time system performance, with
high impact on real-time systems, such as industrial control systems
(ICS).

Our run-time security monitor falls in Class 4, because it (a) models
normal (good) behavior of the system through a formal specification description 
and (b) raises an alarm when the behavior of the application's
execution deviates from the behavior described in the (executable) specification. Specifically, our security monitor has an active model of normal behavior, namely an executable specification of the application~\cite{Shrobe:2006}. This executable specification consists of a decomposition into sub-modules and pre- and post-conditions and invariant for each sub-module. In addition, data-flow and control-flow links connect the sub-modules, specifying the expected flow of values and of control. The pre- and post-conditions and invariant are arbitrary first-order statements about the set of data values (that flow into and out of the sub-modules) and about other arbitrary constraints respectively. 

Our run-time security monitor is suitable not only for new systems, which derive application implementation from application specification, but also for "legacy" systems, where application implementations exist without adequate (formal or informal) application specifications. This can be achieved by describing application specification at any feasible level of abstraction through available specification information. Furthermore, modular application specification at any desired level of abstraction also allows us to monitor only attack(s) specific behavior of "real-time" systems without affecting their performance at run-time.
As our run-time security monitor is using an executable application specification, it is efficient for use in real-time system as has been proven for real-time safety-critical systems~\cite{Doron:2004}.

\begin{figure}
\includegraphics[scale=0.6]{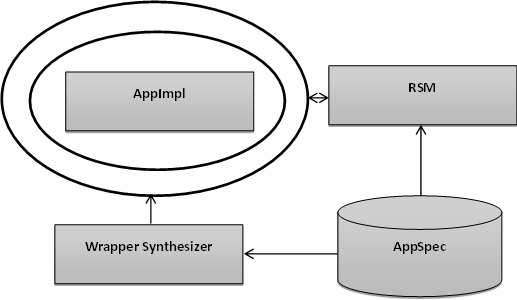}
\caption{The Architecture of Core Defending-System}
\label{fig:monitor}
\end{figure}

Our run-time security monitor (``RSM''), shown in Figure~\ref{fig:monitor}, 
is the core component of a larger system named ARMET. 
ARMET takes as input a specification (``AppSpec'') and an implementation ("AppImpl") of the application of interest.
%that describes the intended behavior of the application of interest. 
Based on the specification, the ``Wrapper Synthesizer'' of ARMET generates probes to observe the run-time  behavior of the application that corresponds to the specification elements. During execution of the ``AppImpl'', the RSM checks whether the actual behavior of the system (\emph{observations} generated by "Wrapper Synthesizer") is consistent with the \emph{predictions} generated from "AppSpec". If an inconsistency is detected, RSM raises an alarm and ARMET suspends the application execution and proceeds to diagnosis, in order to identify why the execution of "AppImpl" did not behave as predicted.
In addition to run-time monitoring, ARMET employs diagnostic reasoning techniques to further isolate and characterize the failure~\cite{Shrobe:1979}. ARMET is highly robust and has high diagnostic data resolution, which is a key requirement of real-time systems that require continuous operation even after a successful attack. ARMET achieves continuous operation through the construction of a far more complex models of applications. 

%Our monitor is suitable not only for new systems like ARMET which derives "AppImpl" from "AppSpec" but also for "legacy" systems where application implementations exist without adequate (formal/informal) application specifications. This can be achieved by describing "AppSpec" at any level of abstraction feasible through available specification information. 

%The task of run-time security monitoring of ``legacy'' and "real-time" systems is challenging because the lack of design information may result in a weak system model and thus the monitor may allow illicit behaviors on one hand and overhead of observing irrelevant system behavior may affect performance of the system at run-time on the other hand. 

%Thus our modeling language addresses above challenges by allowing to specify system behavior at any desired (and pragmatic) level of abstraction and modularity. Moreover, our run-time monitor allows to balance the trade-off between security and performance of the system by observing only required behavior for security at run-time.

%Moreover, model also allows to specify hypothetical attack plans for the components, which are later used by diagnosis process.

RSM runs executable application specification in parallel with the actual application code, comparing their results at the granularity and abstraction level of the executable specification. The executable specification is hierarchical and modular, allowing flexibility in the granularity of the monitoring. Depending on the environment, the executable specification may run at a high level of abstraction, incurring less overhead, but requiring more diagnostic reasoning when the program diverges from the behavior of the executable specification. Alternatively, the executable specification can be elaborated in greater detail, incurring more overhead, but providing more containment.

Optionally, the model can also specify suspected incorrect behaviors of a component and associated potential attack plans, allowing the diagnostic reasoning to characterize the way in which a component may have misbehaved. Then, diagnosis is a selection of behavioral modes for each component of the specification, such that the specification predicts the observed misbehavior of the system.

\enlargethispage*{0.5cm}

Through this work, we introduce a highly reliable run-time security monitor with proven absence of false alarms (i.e. soundness and completeness). Importantly, the proof establishes a contract between the monitor and its user such that, if the user establishes the \emph{assumptions} of the proof, the monitor \emph{guarantees} to detect any violation at run-time.

The remaining of the paper is organized as follows. In Section~\ref{sec:rw}, we describe related work and in Section~\ref{sec:lang} we present the calculus (syntax and semantics) of the application specification language. In Section~\ref{sec:ad}, we first present the calculus of the security monitor and then we present the formulation and proof of soundness and completeness of the monitor. We conclude in  Section~\ref{sec:conc}.
\section{Related Work}\label{sec:rw}
The operation of RSM is to check the consistency between the specified and execution behaviors of an application at run-time. This may be viewed as a run-time verification problem. The goal of run-time verification is to specify the intended behavior of a system in some formalism and to generate an executable monitor from this formalism (i.e. specification) that reports inconsistent execution, if detected.

%In the case of legacy systems, the observed system may specify sufficient and optionally necessary behavior and security properties. Based on the details of specification, monitor may observe interface (pre- and post-conditions) and optionally internal state (invariant) behaviors by which complex properties of the system can be expressed and monitored.

There has been extensive research on specification based run-time monitoring. Most such approaches employ formalism such as context grammars, regular expressions~\cite{Chen:2007}, event calculus~\cite{George:2009}, temporal logic~\cite{Bauer:2011, Barringer:2004} and rule systems operating over atomic formulas~\cite{Barringer:2010}. Such formalism offer limited expressive power to formalize complex system properties, although they can be translated into efficient executable monitors. To addresses the challenges of run-time monitoring of "legacy" and "real-time'' systems (namely the lack of design information and performance respectively), our formalism allows not only to specify dependencies, system level behavior and security properties (in case of partial design details), but
also to specify internal system behavior and complex security properties (in case of desired design details) of such systems as well.

%Based on our proof, we also ensure that the monitor compares the behaviors in a systematic way and thus captures all illicit behaviors.

Run-time monitoring of legacy systems has not received significant attention. However, there have been attempts to apply similar monitoring techniques. 
For example, Kaiser et al. instrument the systems by probing and passing data to another component that forms a basis of the system's model which is later used to monitor run-time modifications automatically~\cite{Kaiser:2010}. More recently, Wofgang et al. have automatically generated run-time monitor for network traffic from a high-level specification language which is based on first order predicate logic and set theory~ \cite{Wolfgang:2015}. Furthermore, based on a variant of denotational semantics of the specification language and operational semantics of the monitor~\cite{Wolfgang:2012}, they verified soundness of the resource analysis of the monitor~\cite{Wolfgang:2014}. The resource analysis identifies the number of instances of the monitor and the number of messages required to detect a violation. 

Model-based executable specifications have been rarely used for run-time monitoring of real-time systems~\cite{Wasserman:1997}. However, Barnett et al. have used ASML as an executable specification language for run-time monitoring~\cite{Barnett:2003}. ASML is an extension of ASM, which is based on the formalism of a transition system whose states are first order algebras~\cite{ASM:2003}. There is no formal semantics of ASML, however, the operational semantics of some constructs of ASM has been defined by Hannan et al.~\cite{Hannan:1992}. More recently, Choilko et al. have developed a framework for executable specification based run-time monitoring of timed systems~\cite{Choilko:2013}. In this work, the formalism of the specification is based on an \emph{extended time interval} which is a pair of a time event and a time interval. The formalism for implementation is based on \emph{timed word} which is a sequence of time events and the goal of the monitor is to check the conformance of an implementation word and the specification trace.

In contrast to the approaches discussed above, the focus of our run-time security monitor is to check consistency of automatically generated \emph{predictions} (conditions) from an executable specification language and run-time \emph{observations} of application execution. The formalism of our specification language is based on monadic second order logic~\cite{Henriksen:1995} and event calculus interpreted over algebraic data structures. This formalism allows specification of faulty behaviors of a system. Furthermore, the formalism enables description of attack plans, which are exploited by the monitor at run-time for early threat detection against more sophisticated and complex attacks, e.g. advanced persistent threats. Our formalism is similar to Crash Hoare-logic that is used to capture the faulty behavior of a file system~\cite{Chen:2015}. Our formalism allows sound construction (resp. specification) of high-level abstract behavior of a system from low-level abstract behavior(s) using a method analogous to classical set builder. Our security monitor is the first approach in run-time monitoring that formally assures the absence of false alarms and thus is sound and complete. For our proof we use the denotational semantics of the application specification language as described in~\cite{MTK:2015a}.

%In detail, the original work~\cite{Shrobe:2006} informally described the security monitor and its modeling language used in the AWDRAT system as a state machine that advances as events are reported from the application.  Here we formalize that description, providing a semantics for the security monitor and its modeling language, which in principle captures their core functional implementation details adequately. Based on this semantics, we show that the formally defined system is sound and complete.

\section{Application Specification Language}\label{sec:lang}
Our executable (application) specification language~\cite{Shrobe:2006} consists of a \emph{decomposition} of an application behavior into sub-modules and pre- and post-conditions and invariant (\emph{behavioral description}) for each sub-module: in rest of the paper, we use the term system for application behavior. The \emph{decomposition} is further equipped with data-flow and control-flow links that connect the sub-modules, specifying the expected flow of values and of control. The specification also allows to specify potential \emph{attack plan}s for the components based on attack models and associated rules that imply a certain attack model.

In the following subsection, we discuss selected high level syntactic domains and their semantics. 
\subsection{Syntax}\label{subsec:sam-syntax}
Based on the aforementioned description, syntactically, the specification language (represented by syntactic domain $\omega$) has following three main top level constructs:
\begin{enumerate}
\item hierarchical decomposition ($\zeta$) of sub-modules,
\item behavioral description ($\eta$) of each sub-module and
\item attack plans ($\epsilon$) of modules/sub-modules.
\end{enumerate}
The simplified grammar of these top level domains is shown in Figure~\ref{fig:domains}.

\begin{figure}[ht]
\centering
%\begin{tabbing}
Application Specification \hspace*{0.9cm} $\omega$ ::= ... $\zeta$ $\eta$ $\epsilon$...
%\\\hspace*{2.8cm}...
\\\hspace*{-0.1cm} Decomposition \hspace*{2.2cm} $\zeta$ ::= $\alpha$ $|$ ($\alpha$) $\zeta$
%\\\hspace*{2.8cm}...
\\Behavioral Model \hspace*{1.8cm} $\eta$ ::= $\beta$ $|$ ($\beta$) $\eta$
\\\hspace*{0.3cm} Attack Plan \hspace*{2.6cm} $\epsilon$ ::= $\delta$ $\rho$  $|$ ($\delta$ $\rho$) $\epsilon$\\
\hspace*{2.8cm}...
%\end{tabbing}
\caption{Top Level Syntactic Domains of the Language}
\label{fig:domains}
\end{figure}

In the following we briefly discuss the decomposition and attack plans, and will focus more on behavioral description, being core and the only one that is also used in the following sections for semantics and proof.
\subsubsection*{Decomposition ($\alpha$)}
The hierarchical decomposition $\alpha$ of a component\footnote{The "component" and "module/sub-module" are used interchangeably.} consists of 
\begin{enumerate}
\item its interface
\begin{itemize}
\item sets of inputs and outputs respectively
%\item a set of its outputs
\item a set of the resources used by the component (e.g. files, binary code, ports) and a set of sub-components
%\item a list of the possible behavior modes that it might exhibit (e.g. normal and one or more failure modes)
%\item a set of sub-components
\item sets of events that allow entry and exit to and from the component respectively
%\item a set of events that allow exit from the component
\item a set of events that are allowed to occur during the execution of the component
\item a set of conditional probabilities between the possible modes of the resources and the possible modes of the component and
a set of known vulnerabilities occurred to the component
\end{itemize}
\item and a structural model that is a set of sub-components some of that might be splits or joins of
\begin{itemize}
\item data-flows between linking ports of the sub-components and
% (outputs of one to inputs of another)
\item control-flow links between cases of a branch and a component that will be enabled if that branch is taken
\end{itemize}
\end{enumerate}
The syntactical domain $\alpha$ is defined in Figure~\ref{fig:decomp}.

\begin{figure}[ht]
\begin{tabbing}
$\alpha$ ::= \textbf{define-}\=\textbf{ensemble} CompName
\\\> \textbf{:entry-events} \hspace*{1cm} \textbf{:auto} $|$ \texttt{set}(Evnt)
\\\> \textbf{:exit-events} \hspace*{1.3cm} \texttt{set}(Evnt)
\\\> \textbf{:allowable-events} \hspace*{0.4cm} \texttt{set}(Evnt)
\\\> \textbf{:inputs} \hspace*{1.95cm} \texttt{set}(ObjName)
\\\> \textbf{:outputs} \hspace*{1.75cm} \texttt{set}(ObjName)
\\\> \textbf{:components} \hspace*{1.1cm} \texttt{set}(Comp)
\\\> \textbf{:controlflows} \hspace*{1.05cm} \texttt{set}(CtrlFlow)
\\\> \textbf{:splits} \hspace*{2.1cm} \texttt{set}(SpltCF)
\\\> \textbf{:joins} \hspace*{2.15cm} \texttt{set}(JoinCF)
\\\> \textbf{:dataflows} \hspace*{1.5cm} \texttt{set}(DataFlow)
\\\> \textbf{:resources} \hspace*{1.55cm} \texttt{set}(Res)
\\\> \textbf{:resource-mapping} \hspace*{0.25cm} \texttt{set}(ResMap)
\\\> \textbf{:model-mappings} \hspace*{0.5cm} \texttt{set}(ModMap)
\\\> \textbf{:vulnerabilities} \hspace*{0.8cm} \texttt{set}(Vulnrablty)
\end{tabbing}
\caption{Syntactic Domain for Decomposition ($\alpha$)}
\label{fig:decomp}
\end{figure}
The elements of $\alpha$ are informally discussed above. Further details of $\alpha$ are out of the scope of this paper.
\subsubsection*{Behavioral Description ($\beta$)}
The $\beta$ describes normal (and optionally various compromised) behavior of a component that includes
\begin{itemize}
\item  set of inputs and outputs respectively,
\item  allowable events during the execution in that mode and
\item  preconditions on the inputs, post-conditions and invariant, all of that are first order logical expressions.
\end{itemize}
The complete syntax of $\beta$ is defined in Figure~\ref{fig:behave}.
\begin{figure}
\begin{tabbing}
$\beta$ ::= \textbf{defbehavior-model} \=(CompName \textbf{normal} $|$ \textbf{compromised})
\\\> \textbf{:inputs} \hspace*{1.5cm} \texttt{set}(ObjName)
\\\> \textbf{:outputs} \hspace*{1.3cm} \texttt{set}(ObjName)
\\\> \textbf{:allowable-events}  \hspace*{0.1cm} \texttt{set}(Evnt)
\\\> \textbf{:prerequisites} \hspace*{0.6cm} \texttt{set}(BehCond)
\\\> \textbf{:postconditions} \hspace*{0.4cm} \texttt{set}(BehCond)
\\\> \textbf{:invariant} \hspace*{1.1cm} \texttt{set}(BehCond)
\end{tabbing}
\caption{Syntactic Domain for Behavioral Description ($\beta$)}
\label{fig:behave}
\end{figure}

\subsubsection*{Attack Plan ($\epsilon$)}
The attack plan $\epsilon$ consists of a description of potential attack models ($\delta$) and the rules ($\rho$) that imply a certain attack.  Syntactically, an attack plan includes
\begin{itemize}
\item a set of types of attacks that are being anticipated and the prior probability of each of them,
\item a set of effects such that how each attack type can effect mode (normal/compromised) of a resource and
\item a set of rules expressing the conditional probabilities between attack types and resource modes.
\end{itemize}

The syntactic domains of $\delta$ and $\rho$ are defined in Figure~\ref{fig:attack} resp.

\begin{figure}
\begin{tabbing}
$\delta$ ::= \textbf{define-}\=\textbf{attack-model} AtkModName
\\\> \textbf{:attack-types} \hspace*{1.5cm} (\texttt{set}(AtkType))
\\\> \textbf{:vulnerability-mapping} \hspace*{0.1cm} (\texttt{set}(AtkVulnrabltyMap))
\end{tabbing}

\begin{tabbing}
$\rho$ ::= \textbf{defrule} \=AtkRulName (\textbf{:forward})
\\\> \textbf{if} \texttt{set}(AtkCond) \textbf{then} \texttt{set}(AtkCons)
\end{tabbing}
\caption{Syntactic Domains of Attack Model ($\delta$) and Rule ($\rho$)}
\label{fig:attack}
\end{figure}

In principle, attack plans are hypothetical attacks based on rules that describe different ways of compromising a component. The monitor exploits such plans to match at run-time and detect any such attack, thus making the monitor more robust.

%In the following section, we explain the language constructs with the help of an example.
\subsection{Example}\label{subsec:example}
\begin{figure}
\centering
\includegraphics[scale=0.4]{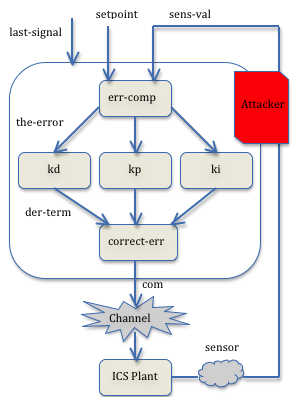}
\includegraphics[scale=0.5]{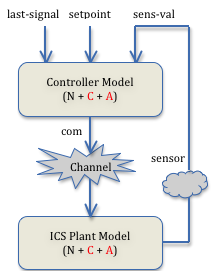}
(a) Application \hspace*{2cm} (b) Specification
\caption{A Controller Application and its Model}
\label{fig:example}
\end{figure}

To provide an intuitive grounding for these ideas we will consider an example of a simple ICS and of its model in the specification language.  The system consists of a water tank, a level sensor and a pump that is capable of either filling or draining the tank. The tank has a natural leakage rate that is proportional to the height of the water column in the tank.
The tank is controlled by a PID controller; this is a computational device running a standard (PID) control algorithm that has a simple structure:

The algorithm has two  inputs: The {\em set-point}, i.e. the water level that the tank should maintain and the {\em sensor value} provide by the level sensor.  It has a simple output, the {\em command}.
The algorithm performs the following computations based on the three parameters notated as {\em Kp}, {\em Ki} and {\em Kd} that are used as scaling weights in the algorithm as shown in Figure~\ref{fig:example} (a).

\begin{enumerate}
\item Calculate the {\em error}, the difference between the set-point and the sensor value
\item Calculate three terms:
 \begin{enumerate}
    \item The {\em Proportional} term; this is just the error weighted by Kp.
    \item  The {\em Integral} term; this is a running sum of the {\em errors} seen so far, weighted by Ki.
    \item The {\em Derivative} term; this is a local estimate of rate of change of the sensor value, weighted by Kd.
 \end {enumerate}
\item Calculate the sum of the three terms.
\item The value of the sum is the {\em command} output of the algorithm.
\end {enumerate}

The {\em command} output of the algorithm is sent to the pump, controlling the rate at which the pump either adds or removes water.  The algorithm is ``tuned'' by the choice of the three parameters Kp, Ki  and Kd; when well tuned the system responds quickly to deviations from the set-point with little over-shoot and very small oscillations around the set-point. 
%The system consisting of the pumps, tank, sensor and controller is a closed-loop system that never terminates.

Finally, we note that the level sensor can be viewed as (and often is) a computational and communication device that estimates the actual height of the water tank and communicates the estimated height back to the controller. 

There are two standard categories of attacks on such a system: 
\begin{itemize}
\item {\bf False Data Injection Attacks}.  These are attacks on the sensor and its communication channel, such that the controller receives a value that is different from the actual level of the tank.
\item {\bf Controller Attacks}.  These are penetrations to the computer running the control algorithm. For our purposes it is only necessary to consider attacks that overwrite the value of one of Kp, Ki, or Kd.  Any such attack, will cause the controller to calculate an incorrect command.
\end{itemize}

In either case, the end result is that the level in the water tank will not be correctly maintained. In the first case, the controller calculates a correct response to the distorted sensor value.  For example, suppose that the attacker is systematically distorting the sensor value to be too low. In that case, the controller will continuously issue commands to the pump to add water to the tank, eventually causing the tank to overflow. In the second case, a change in value of one of the controller parameters will cause the controller to calculate in an incorrect command. This can have a variety of effects, depending on which parameters are changed.  
%An extreme example involves zeroing out Kd and Ki and changing the sign of Kp; this would cause the controller to add water to the tank when the water level is already too high and to drain water when the level is too low.

Monitoring of such a system requires its behavioral specification as shown in Figure~\ref{fig:example} (b). The actual system is a {\em cyber-physical} system, containing both physical components (i.e. the tank, the pump) and computational components (i.e. the controller and the sensor). The monitor model parallels this structure; it contains computational models of the controller and the sensor as well as a computational model of the physical plant. This later model performs a numerical integration of the differential equations describing the physical plant's behavior, e.g. the dynamics of the pump. 
%In this case, the model is quite simple, but more complicated systems would require more complex models.    
The application specification of the controller, essentially mirrors the structure of the algorithm: There is a component that calculates the error term, data-flow links that connect the error term to each of three parallel steps that calculate the Proporational, Integral and Derivative terms, finally there is the summation component that adds the three terms, calculating the command output.

%The models for these components are straightforward.  For example the normal behavioral model for the Kd calculation simply states that the computation happened correctly:
%Whereas the abnormal behavioral model simply states that the output value isn't the expected result as shown in Figure~\ref{fig:behave-comp-der-example}. Here, runtime behavior of the monitor will depend on strength of the predicate \texttt{isgoodsignal}, e.g. if the predicate is too weak, the monitor may allow undesired behaviors. In the later sections, we show that the monitor only captures any undesired behavior.
%Furthermore, model of decomposition of the controller is shown in Figure~\ref{fig:decomp-controller-example} where N and C refers to normal and compromised behavior and A refers to possible attack model. 

The structural model of the controller is shown diagrammatically in
Figure~\ref{fig:example} (b) (N and C refers to normal and compromised behavior and A
refers to possible attacks).  The models for the components of the
controller are reasonably straightforward. For example, the normal
behavioral model for the Kd calculation states that the output of the
component is the derivative of the error, weighted by Kd.  This is
expressed as a post-condition, as shown in Figure~\ref{fig:behave-comp-der-example}.

Notice that what the controller calculates is a discrete approximation of the derivative of
the error term, which is calculated using the previous and current
versions of the error.  The value of the error term is conceptually a
state variable that is updated between successive iterations of
the controller computation.  In our specification language, however,
we model these as extra inputs and data flows  (as we do also for control algorithm parameters
such as Kd).  For simplicity, we have omitted these extra items from the diagram
in Figure~\ref{fig:example}.

The compromised behavioral model states that any other
behavior is acceptable; it does so by stating no post-conditions.

The run-time behavior of the monitor will depend
on the strength of the post-conditions; if these are
too weak, the monitor may allow undesired behaviors..

\begin{figure}
{\small
\begin{verbatim}
(define-component-type controller-step
 :entry-events (controller-step)
 :exit-events (controller-step)
 :allowable-events (update-state accum-error)
 :inputs (set-point sens-val)
 :outputs (com)

  :components 
((err-comp :type err-comp :models (normal))
(comp-der :type comp-der :models (normal)) ... )

  :dataflows 
((set-point controller-step set-point err-comp)
(the-error err-comp the-error comp-der)...))
\end{verbatim}
}
\caption{Decomposition of the Module \texttt{controller-step}}
\label{fig:decomp-controller-example}
\end{figure}

%\begin{figure}
%\begin{verbatim}
%(define-component-type estimate-error
%  :entry-events (estimate-error)
%  :exit-events (estimate-error)
%  :inputs (observation)
%  :outputs (the-error)
%  :behavior-modes (normal)
%  )
%
%(defbehavior-model (estimate-error normal)
%  :inputs (observation)
%  :outputs (the-error)
%  :prerequisites ([data-type-of ?observation number])
%  :post-conditions ([data-type-of ?observation number])
%  )
%\end{verbatim}
%\caption{Behavior of the Module \texttt{estimate-error}}
%\end{figure}
\begin{figure}
{\small
\begin{verbatim}
(define-component-type comp-der
 :entry-events (compute-derivative)
 :exit-events (compute-derivative)
 :inputs (the-error old-error kd time-step)
 :outputs (der-term)
 :behavior-modes (normal compromised) )

(defbehavior-model (comp-der normal)
 :inputs (the-error the-old-error kd time-step)
 :outputs (der-term)
 :prerequisites ([data-type-of the-error number])
 :post-conditions 
 	([and [data-type-of der-term number]
   [equal der-term 
   	(*kd(/(- new-error old-error) time-step))]]))

 (defbehavior-model (comp-der compromised)
 :inputs (the-error the-old-error kd time-step)
 :outputs (der-term)
 :prerequisites ()
 :post-conditions ())
\end{verbatim}
}
\caption{Normal and Compromised Behavior of \texttt{comp-der}(kd)}
\label{fig:behave-comp-der-example}
\end{figure}

%\enlargethispage*{2.5cm}
%\begin{figure}
%\centering
%\includegraphics[scale=0.4]{example-model.png}
%\caption{An Example Controller Model}
%\label{fig:model}
%\end{figure}

%In the following, we use behavior element $\beta$ of the specification language to define its formal semantics of the specification language and proofs of soundness and completeness are based on the  that are discussed in the following subsections.
%In the following subsection, we discuss the formal semantics of the description of behavioral model.
\subsection{Formal Semantics}\label{subsec:sam-semantics}
In this section, we first give the definition of semantic algebras, then discuss informal description and the formal denotational semantics of the core construct (i.e. behavioral description) of the specification language.
\subsubsection{Semantic Algebras}
\emph{Semantic domains}\footnote[1]{These domains are common to a program to be monitored, its specification language and the monitor.} \footnote[2]{We use subscript $s$ and $r$ to specify domains for specification and program's runtime resp., e.g. State$_s$ = specification state, State$_r$ = program's runtime state, State = combined monitor state.} represent a set of elements that share some common properties. A semantic domain is accompanied by a set of operations as functions over the domain. A domain and its operations together form a \emph{semantic algebra}~\cite{Schmidt86}.  The domains of our language are similar to the domains of any classical programming/specification language (e.g. Java, JML, ACSL). In the following we declare/define only important semantic domains and their operations.

\subsubsection*{Environment Values}
The domain \emph{Environment} holds the environment values of the language and is formalized as a tuple of domains \emph{Context} (which is a mapping of identifiers to the environment values) and \emph{Space} (that models the memory space). The \emph{Environment} domain includes interesting values, e.g. component, attack plan and resource. Here resource can be binary code in memory, files and ports etc.

\noindent\textbf{Domain}: \emph{Environment}\\
\emph{Environment} := \emph{Context} $\times$ \emph{Space}\\
\emph{Context} := \emph{Identifier} $\rightarrow$ \emph{EnvValue}
\\\emph{EnvValue} := \emph{Variable} + \emph{Component} + \emph{AtkPlan} + \emph{Resource} + ...
\\\emph{Space} := $\mathbb{P}$(\emph{Variable})\\
\emph{Variable} := n, where n $\in$ $\mathbb{N}$ represents locations
%\\\textbf{Operations}:
%\begin{itemize}
%\item space: Environment $\rightarrow$ Space
%\item context: Environment $\rightarrow$ Context
%\item environment: Context $\times$ Space $\rightarrow$ Environment
%\item take: Space $\rightarrow$ Identifier $\times$ Space\\
%take(s) = LET x = SUCH x: x $\in$ s IN \textless x, s$\setminus$\{x\}\textgreater
%\item push: Environment $\times$ Identifier $\times$ Component $\rightarrow$ Environment
%\begin{tabbing}
%push(e, I, c) = LET \=\textless x, s'\textgreater = take(space(e)) IN \\\>environment(context(e)[I $\mapsto$ inComponent(c)], s')
%\end{tabbing}
%\end{itemize}

The domain \emph{Environment} supports typical selection, update and equality operations over its values.

\subsubsection*{State Values}
The domain \emph{State} represents the execution of a program. A \emph{Store} is important element of the state and holds for every \emph{Variable} a \emph{Value}. The \emph{Data} of the state is a tuple of a \emph{Flag} that represents the current status of the state and a \emph{Mode} to represent the current mode of execution of the state of a component.

\noindent\textbf{Domain}: \emph{State}\\
\emph{State} := \emph{Store} $\times$ \emph{Data}\\
\emph{Store} := \emph{Variable} $\rightarrow$ \emph{Value}\\
\emph{Data} := \emph{Flag} $\times$ \emph{Mode}\\
\emph{Flag} := \{running, ready, completed\}\\
\emph{Mode} := \{normal, compromised\}
%\\\textbf{Operations}:
%\begin{itemize}
%%\item state: Store $\times$ Flag $\rightarrow$ State\\
%%state(s,f) = \textless s,f\textgreater
%%\item store: State $\rightarrow$ Store\\
%%store(\textless s,f\textgreater) = s
%%\item data: State $\rightarrow$ Data\\
%%data(\textless s,d\textgreater) = d
%%\item flag: Data $\rightarrow$ Flag\\
%%flag(\textless f,m\textgreater) = f
%%\item mode: Data $\rightarrow$ Mode\\
%%mode(\textless f,m\textgreater) = m
%\item setFlag: State $\times$ Flag $\rightarrow$ State
%%setFlag(s, f) = LET d = \textless f, mode(data(s))\textgreater IN \textless s, d\textgreater
%\item eqFlag: State $\times$ Flag $\rightarrow$ Bool
%%eqFlag(s, f) = IF equals(flag(data(s)), f) THEN \texttt{true} ELSE \texttt{false} END
%\item setMode: State $\times$ Mode $\rightarrow$ State
%%setMode(s, m) = LET d = \textless flag(data(s)), m\textgreater IN \textless s, d\textgreater
%\item eqMode: State $\times$ Mode $\rightarrow$ Bool
%%eqMode(s, m) = IF equals(mode(data(s)), m) THEN \texttt{true} ELSE \texttt{false} END
%\item update: State $\times$ Variable $\times$ Value $\rightarrow$ State
%%update(s, var, val) = state(store(s)[var $\mapsto$ val], flag(s))
%\end{itemize}

The domain \emph{State} has typical operations, e.g. read and write/update of values, checking equality of \emph{Flag} and \emph{Mode} in a given state, and setting a certain \emph{Flag} and \emph{Mode} of a given state. 

\subsubsection*{Semantic Values}
\emph{Value} is a disjunctive union domain and note that the domain \emph{Value} is a recursive domain.

\noindent\textbf{Domain}: \emph{Value}
\begin{tabbing}
\emph{Value} := \emph{ObsEvent} + \emph{RTEvent} + \emph{Component} + \emph{AtkPlan} + ... + \emph{Value}$^*$
\end{tabbing}
%\textbf{Operations}:
%\begin{itemize}
%\item equals: Value $\times$ Value $\rightarrow$ Bool
%\end{itemize}
The domain includes semantic values of observable event, a run-time event and attack plan etc.
The equality of the given two semantic values can be evaluated.
\subsubsection*{Component Values}
The \emph{Component} formalizes the semantic model of a component as a predicate over decomposition, normal and compromised behavior and a pre-state and a post-state of the component's execution respectively. The predicate is formalized as follows:
\begin{center}
\emph{Component} = $\mathbb{P}$(\emph{SBehavior} $\times$ \emph{NBehavior} $\times$ \emph{CBehavior} $\times$ \emph{State} $\times$ \emph{State}$_\bot$\footnote[3]{\emph{State}$_\bot$ = \emph{State} $\cup$ \{$\bot$\}})
\end{center}
where
\begin{tabbing}
\emph{SBehavior} := $\mathbb{P}$(\emph{Value}$^*$ $\times$ \emph{Value}$^*$ $\times$ \emph{Value}$^*$ $\times$ \emph{State} $\times$ \emph{State}$_\bot$)\\
\emph{NBehavior} = \emph{CBehavior} := $\mathbb{P}$(\emph{Value}$^*$ $\times$ \emph{Value}$^*$ $\times$ \emph{State} $\times$ \emph{State}$_\bot$)
\end{tabbing}
Furthermore, \emph{SBehavior} is defined as a predicate over sets of input and output values, set of allowable values, a pre-state and a post-state of the behavior. Also, normal behavior and compromised behavior (\emph{NBehavior} and \emph{CBehavior}) are also defined as predicates over sets of input and output values, a pre-state and a corresponding post-state respectively. 

\subsubsection*{Attack Values}
The semantics domain \emph{AtkModel} formalizes the attack model and is defined as a predicate over an attack name, probability of the attack and the corresponding vulnerability causing the attack; the attack model is formulated as follows:
\begin{center}
\emph{AtkModel} := $\mathbb{P}$(\emph{Identifier} $\times$ \emph{FVal} $\times$ \emph{Vulnerability})
\end{center}
%Note that the two predicates are the valuation functions of corresponding syntactic domains.

\subsubsection{Signatures of Valuation Functions}
A valuation function defines a mapping of a language's abstract syntax structures to its corresponding meanings (semantic algebras)~\cite{Schmidt86}. The valuation function operates on a syntactic construct and returns a function from the environment to a semantic domain.

%A valuation function VF for a syntax domain VF is usually formalized by a set of equations, one per alternative in the corresponding BNF for each syntactic domain.

We define the result of the valuation function as a predicate, e.g. the behavioral relation (BehRelation) is defined as a predicate over an environment, a pre- and a post-state and is defined as follows:
\begin{center}
BehRelation := $\mathbb{P}$(Environment $\times$ State $\times$ State$_\bot$)
\end{center}

The valuation functions for the abstract syntax domains of specification ($\omega$), behavioral description ($\beta$) and attack plans ($\epsilon$) have same signatures. For example, a valuation function signature for $\beta$ is defined as follows:
%\subsubsection*{System Architectural Model}
\begin{center}
%\textlbrackdbl $\omega$\textrbrackdbl : Environment $\rightarrow$ BehRelation
\textlbrackdbl $\beta$\textrbrackdbl : Environment $\rightarrow$ BehRelation
%\\\textlbrackdbl $\epsilon$\textrbrackdbl : Environment $\rightarrow$ BehRelation
\end{center}

%%\subsubsection*{Behavioral Models}
%The signature of valuation function for the behavioral model:
%\begin{center}
%\textlbrackdbl $\beta$\textrbrackdbl : Environment $\rightarrow$ BehRelation
%\end{center}
\enlargethispage*{1cm}
Based on the above relation and the auxiliary semantic inference rules (see Figure~\ref{fig:semrules}), we define valuation functions for $\beta$ and $\epsilon$ in the following subsection.
\subsubsection{Definition of Valuation Functions}
Semantically, normal and compromised behavioral models results in modifying the corresponding elements of the environment value \emph{Component} as defined below:
\begin{tabbing}
\textlbrackdbl\=$\beta$\textrbrackdbl (e)(e', s, s') $\Leftrightarrow$ 
\\\> LET \= c $\in$ Component: \textlbrackdbl CompName\textrbrackdbl(e)(s, s', inValue(c)) IN
\\\> $\forall$ \=e$_1$ $\in$ Environment, nseq $\in$ \texttt{set}(EvntName), b$_1$, b$_2$: $\mathbb{B}$,
\\\>\> eseq $\in$ ObsEvent*, iseq, oseq $\in$ Value$^*$: \\\>\> \textlbrackdbl \texttt{set}(ObjName$_1$)\textrbrackdbl (e)(s, iseq) $\wedge$ \textlbrackdbl \texttt{set}(BehCond$_1$)\textrbrackdbl (e) (s) $\wedge$
\\\>\> {\tt noatk}(c, e, b$_1$) $\wedge$ \textlbrackdbl \texttt{set}(Evnt)\textrbrackdbl (e) (e', s, s', nseq, eseq) $\wedge$
\\\>\> \textlbrackdbl \texttt{set}(ObjName$_2$)\textrbrackdbl (e')(s', oseq) $\wedge$\textlbrackdbl \texttt{set}(BehCond$_2$)\textrbrackdbl(e')(s,s')$\wedge$ 
\\\>\> \textlbrackdbl \texttt{set}(BehCond$_3$)\textrbrackdbl (e') (s, s') $\wedge$ {\tt noatk}(c, e', b$_2$)
\\\>\>$\Rightarrow$
\\\>\> LET \=v = b$_1$ $\wedge$ b$_2$ $\wedge$ eqMode(s', "normal") IN
\\\>\>\> {\tt update}(c, e', s, s', iseq, oseq, v)
%\\\>\> IF \=eqMode(inState$_\bot$(s'), ``normal'') THEN
%\\\>\>\> LET \= sbeh = c[1], nbeh = \textless inseq, outseq, s, s'\textgreater, cbeh = c[3] IN 
%\\\>\>\>\> e' = push(e$_1$\=, store(inState(s'))(\textlbrackdbl CompName\textrbrackdbl(e$_1$))
%\\\>\>\>\>\>, c(sbeh, nbeh, cbeh, s, s'))
%\\\>\>\> END
%\\\>\> ELSE
%\\\>\>\> LET \= sbeh = c[1], nbeh = c[2], cbeh = $\langle$inseq, outseq, s, s'$\rangle$~ IN 
%\\\>\>\>\> e' = push(e$_1$\=, store(inState(s'))(\textlbrackdbl CompName\textrbrackdbl(e$_1$))
%\\\>\>\>\>\>, c(sbeh, nbeh, cbeh, s, s'))
%\\\>\>\> END
%\\\>\> END
\end{tabbing}
where {\tt update} is an auxiliary semantic rule as shown in Figure~\ref{fig:semrules}.
\begin{figure}

\begin{mathpar}
\inferrule
  {t \in \{{\tt ENTRY, EXIT, ALLOWABLE, NONE}\}}
  {{\tt typeOf}(oe, c) \rightarrow t}
\end{mathpar}

\begin{mathpar}
\inferrule
  {\text{dataArrives}(c, s(i), s'(i)) \\ {\tt comp}(c, e(i), e'(i), s(i), s'(i), {\texttt{False}}, 0) \\ s(\text{i+1}) = s(i) \\ s'(\text{i+1}) = s(i) \\ \text{setMode}(s'(\text{i+1}), \text{"compromised"})}
  {{\tt run}({\tt ENTRY}, c, e, e', s, s', i, {\tt False})}
\end{mathpar}

\begin{mathpar}
\inferrule
  {\text{dataArrives}(c, s(i), s'(i)) \\ {\tt comp}(c, e(i), e'(i), s(i), s'(i), {\texttt{True}}, 0) \\  \text{setFlag}(s'(\text{i+1}), \text{"running"}) \\ s(\text{i+1}) = s'(i) \\ e(\text{i+1}) = e'(i) \\ cseq = components(c) \\ {\tt mon}(cseq, s(\text{i+1}), s'(\text{i+1}), e(\text{i+1}), e'(\text{i+1}))}
  {{\tt run}({\tt ENTRY}, c, e, e', s, s', i, {\tt True})}
\end{mathpar}

\begin{mathpar}
\inferrule
  {\text{dataArrives}(c, s(i), s'(i)) \\ \text{setFlag}(s'(\text{i+1}), \text{"completed"}) \\ {\tt comp}(c, e(i), e'(i), s(i), s'(i), b, 1) \\ s(\text{i+1}) = s'(i) \\ s'(\text{i+1}) = s'(i) \\ [ b = {\tt False} \Rightarrow \text{setMode}(s'(\text{i+1}), \text{"compromised"}) ] \\ [ b = {\tt True} \Rightarrow \text{setMode}(s'(\text{i+1}), \text{"normal"}) ]}
  {{\tt run}({\tt EXIT}, c, e, e', s, s', i, b)}
\end{mathpar}

%\begin{mathpar}
%\inferrule
%  {\text{dataArrives}(c, s(i), s'(i)) \\ \text{setFlag}(s'(i+1), \text{"completed"}) \\ {\tt comp}(c, e(i), e'(i), s(i), s'(i), {\texttt{True}}, 1)  \\ s(\text{i+1}) = s'(i) \\ e(\text{i+1}) = e'(i) \\ \text{setFlag}(s'(\text{i+1}), \text{"normal"})}
%  {{\tt run}({\tt EXIT}, c, e, e', s, s', i, {\tt True})}
%\end{mathpar}

\begin{mathpar}
\inferrule
  {\text{inv}(c, e(i), e'(i), s(i), s'(i), b_1) \\ \text{noatk}(c, e(i), b_1) \\ s(\text{i+1}) = s'(i) \\ e(\text{i+1}) = e'(\text{i+1})}
  {{\tt run}({\tt ALLOWABLE}, c, e, e', s, s', i, b_1 \wedge b_2)}
\end{mathpar}

\begin{mathpar}
\inferrule
  {\text{setMode}(s'(\text{i}), \text{"compromised"} \\ s(\text{i+1}) = s'(i) \\ e'(\text{i+1}) = e(\text{i+1})}
  {{\tt run}({\tt NONE}, c, e, e', s, s', i, {\tt Flase})}
\end{mathpar}

\begin{mathpar}
\inferrule
  {nbeh=\langle inseq, outseq, s, s'\rangle \\ c' = \langle c[1], nbeh, c[3], s, s'\rangle}
  {{\tt update}(c, e_1[id(c) \mapsto c'], s, s', inseq, outseq, {\texttt{True}})}
\end{mathpar}

\begin{mathpar}
\inferrule
  {cbeh=\langle inseq, outseq, s, s'\rangle \\ c' = \langle c[1], c[2], cbeh, s, s'\rangle}
  {{\tt update}(c, e_1[id(c) \mapsto c'], s, s', inseq, outseq, {\texttt{False}})}
\end{mathpar}

\begin{mathpar}
\inferrule
  {a = \langle aseq, apseq, vnseq\rangle}
  {{\tt atk}(atkName, e, e[atkName \mapsto a], aseq, apseq, vnseq)}
\end{mathpar}

\begin{mathpar}
\inferrule
  {b=[\forall at:\text{AtkName}:at=\text{context}(e)(\text{AtkName}) \Rightarrow \text{notcomp}(c, at)]}
  {{\tt noatk}(c, e, b)}
\end{mathpar}

\begin{mathpar}
\inferrule
  {\text{inv}(c, e(i), e'(i), s(i), s'(i), b_1) \\ b_2 = [ x=0 \Rightarrow \text{precond}(c, e(i), e'(i), s(i), s'(i), {\texttt{True}}) ] \\ b_3 = [ x=1 \Rightarrow \text{postcond}(c, e(i), e'(i), s(i), s'(i), {\texttt{True}}) ] \\ {\tt noatk}(c, e(i), b_4)}
  {{\tt comp}(c, s, s', e, e', b_1 \wedge b_2 \wedge b_3 \wedge b_4, x)}
\end{mathpar}

%\begin{mathpar}
%\inferrule
%  {\text{inv}(c, e(i), e'(i), s(i), s'(i), {\texttt{False}}) \\ \vee \\ x=0 \rightarrow \text{precond}(c, e(i), e'(i), s(i), s'(i), {\texttt{False}}) \\ \vee \\ x=1 \rightarrow \text{post}(c, e(i), e'(i), s(i), s'(i), {\texttt{False}}) \\ \vee \\ \text{noatk}(c, e(i), {\texttt{False}})}
%  {{\tt comp}(c, s, s', e, e', \texttt{False}, x)}
%\end{mathpar}

\begin{mathpar}
\inferrule
  {\exists rte \\ \text{arrives}(rte, s) \\ \text{monitors}(\text{i+1}, rte, c, e, e'', s, s'') \\ {\tt mon}(cseq, s'', s', e'', e', s'', s', i)}
  {{\tt mon}(c;cseq, s, s', e, e', s, s', i)}
\end{mathpar}
\caption{Auxiliary Semantic Inference Rules}
\label{fig:semrules}
\end{figure}

In detail, if the semantics of $\beta$ in an environment $e$ yields environment $e'$ and transforms a pre-state $s$ into a post-state $s'$ then
\begin{itemize}
\item the evaluation of inputs \texttt{set}(ObjName$_1$) yields a set of values $iseq$ in environment $e$ and state $s$ such that the pre-conditions \texttt{set}(BehCond$_1$) hold in $e$ and $s$ and the component $c$ has no potential threat ({see rule \tt noatk}) and
\item the evaluation of allowable events results in environment $e'$ and given post-state $s'$ with some auxiliary sets $nseq$ and $eseq$ and
\item the evaluation of outputs \texttt{set}(ObjName$_2$) yields a set of values $oseq$ in $e'$ and $s'$ such that post-conditions \texttt{set}(BehCond$_2$) hold in $e_1$, $s$ and $s'$ and
\item the invariant  \texttt{set}(BehCond$_2$) holds in $e'$, $s$ and $s'$, and the component $c$ has no threat ({\tt noatk}), finally the environment $e'$ can be constructed as follows
\begin{itemize}
\item if the post-state is ``normal'' then $e'$ is an update to the normal behavior ``nbeh'' of the component ``CompName''
\item otherwise $e'$ is an update to the compromised behavior ``cbeh'' of the component as shown in the corresponding inference rules of {\tt update}.
\end{itemize}
\end{itemize}
%In the construction of the environment $e'$ the rest of the semantics of the component do not change as represented in the corresponding LET-IN constructs.
\enlargethispage*{1cm}
Moreover, the valuation function for attack plan is defined as:
\begin{tabbing}
\textlbrackdbl \=$\delta$\textrbrackdbl (e)(e', s, s') $\Leftrightarrow$ 
\\\> $\forall$ \= s'' $\in$ State, aseq, aseq', vnseq $\in$ ISeq, apseq $\in$ Value$^*$: \\\>\> \textlbrackdbl \texttt{set}(AtkType)\textrbrackdbl (e)(s, inState$_\bot$(s''), aseq, apseq) $\wedge$ 
\\\>\> \textlbrackdbl \texttt{set}(AtkVulnrabltyMap)\textrbrackdbl (e) (s'', s', aseq', vnseq) $\wedge$ 
\\\>\> {\tt atk}(AtkModName, e, e', aseq, apseq, vnseq)
%\\\>\> LET \= amod $\in$ $\delta$ IN 
%\\\>\>\> e' = push(e\=, store(inState(s'))(\textlbrackdbl AtkModName\textrbrackdbl(e)), amod(aseq, apseq, vnseq)))
%\\\>\> END \\respectsOrder(aseq, aseq') $\wedge$ 
\end{tabbing}

In detail, the semantics of the domain ``$\delta$'' updates the environment $e$ with a semantic value of \emph{AtkPlan} such that if
\begin{itemize}
\item in a given $e$ and $s$, the evaluation of ``\texttt{set}(AtkType)'' yields post-state $s''$, a set of attack types $aseq$ and a set of values (conditional probabilities) $apseq$ and also
\item in given $e$ and $s$, the evaluation of ``\texttt{set}(AtkVulnrabltyMap)'' yields post-state $s'$, a set of attack types $aseq'$ and a set of vulnerabilities $vnseq$, then
\item the environment $e'$ is an update of environment $e$ with the semantic value \emph{AtkPlan}, which is a triple of (a) a set of attack types (b) a set of corresponding probabilities and (c) a set of vulnerabilities causing the attack types, respectively.
%\begin{enumerate}
%\item a set of attack types,
%\item a set of corresponding probabilities and
%\item a set of vulnerabilities causing the attack types, respectively.
%\end{enumerate}
\end{itemize}

%However, if the semantics of the syntactic domain ``AtkRule'' holds in an environment $e$, then
%\begin{itemize}
%\item there is some resource $r$ such that (as given in ``AtkCondSeq'' respective ``AtkCond'')
%\begin{enumerate}
%\item the resource name is ``?resource-name''  and
%\item the resource type is ``ResType'' and
%\item if the resource has been compromised by an attack ``AtkTypeName'', then
%\end{enumerate}
%\item the resource $r$ (and its associated component $c$) has behavior as specified by the evaluation of consequences ``AtkCodSeq'' in an environment $e$ and state $s$.
%\end{itemize}
%Formally, the semantics of the syntactic domain ``AtkRule'' is defined as:
%\begin{tabbing}
%\textlbrackdbl \=AtkRule\textrbrackdbl (e)(e', s, s') $\Leftrightarrow$ 
%\\\> $\exists$ \=r $\in$ Resource, c $\in$ Component: \textlbrackdbl AtkCondSeq\textrbrackdbl (e)(s, s', r, c) $\wedge$
%\\\>\> \textlbrackdbl AtkConsSeq\textrbrackdbl (e) (s, s', r, c) $\wedge$ e' = e
%\end{tabbing}

\section{Security Monitor}\label{sec:ad}
Based on~\cite{Shrobe:2006}, in this section we discuss the informal behavior of our run-time security monitor whose main goal is to check consistency between a program's run-time  \emph{observations} and its specification-based \emph{predictions} and to only raise a flag if any inconsistency is identified. In detail, when the application implementation starts execution, a ``startup'' event is generated and
dispatched to the top level component of the system, which transforms the execution state of the component
into ``running'' mode. The component instantiates its subnetwork (i.e. sub-components) and propagates
the data along its data-links by enabling the corresponding control-links (if involved). When the data arrives on the input port of the component, the monitor checks if it is complete; if so, the monitor checks the preconditions of the component for the data and if they succeed, it transforms the state of the component into ``ready'' mode. Should the conditions fail, it raises a flag.

After the above startup, the execution monitor starts monitoring the arrival of every \emph{observation} (run-time event) as follows:
\begin{enumerate}
\item If the event is a ``method entry'', then the execution monitor checks if this is one of the ``entry events'' of the component in the ``ready'' state; if so, then after receiving the data, the respective preconditions, invariant and absence of attack plans are checked; if they succeed, then the data is applied on the input port of the component and the mode of the execution state is changed to ``running''.
\item If the event is a ``method exit'', then the execution monitor checks if this is one of the ``exit events'' of the component in the ``running'' state; if so, it changes its state into ``completed'' mode and collects the data from the output port of the component and checks for the corresponding postconditions, invariant and absence of attack plans. Should the checks fail, the monitor raises an alarm.
\item If the event is one of the ``allowable events'' of the component, if invariant holds and there is no attack plan then it continues execution and finally
\item otherwise, if the event is an none of the above events, then the monitor raises an alarm.
\end{enumerate}

%Based on the above informal description, in the following subsection, we give formal semantics of the monitor.
\subsection{Formal Semantics}\label{subsec:ad-semantics}
Based on the aforementioned description of the execution monitor, we have formalized the denotational semantics of the monitor by a relation $monitor$ that is declare and defined as follows:

\begin{tabbing}
\textbf{monitor}  $\subseteq$ \=\textbf{AppImpl} $\times$ \textbf{AppSpec} 
\\\>$\rightarrow$ \textbf{Environment} $\rightarrow$ \textbf{State} $\times$ \textbf{State}$_\bot$
%\end{tabbing}
%\begin{tabbing}
%\textbf{monitor}  $\subseteq$ \=\textbf{AppImpl} $\times$ \textbf{AppSpec} 
%\\\>$\rightarrow$ \textbf{Environment} $\rightarrow$ \textbf{State} $\times$ \textbf{State}$_\bot$
\\monitor($\kappa$, $\omega$)(e)(s, s') $\Leftrightarrow$
\\$\forall$ \=c $\in$ Component, t, t' $\in$ State$_s$, d, d' $\in$ Environment$_s$, rte $\in$ RTEvent:
\\\> \textlbrackdbl $\omega$\textrbrackdbl(d)(d', t, t') $\wedge$ \textlbrackdbl $\kappa$\textrbrackdbl(e$_r$)(e$_r$', s, s') 
$\wedge$ setFlag(s, ``running'') $\wedge$ 
\\\>eqMode(s, "normal") $\wedge$ arrives(rte, s) $\wedge$ equals(s, t) $\wedge$ equals(e$_r$, d)
\\$\Rightarrow$ \=
\\\>$\forall$ \=p, p' $\in$ Environment*, m, n $\in$ State*: 
\\\>\>equals(m(0), s) $\wedge$ equals(p(0), e$_r$) $\wedge$
\\\>\>$\exists$ \=k $\in$ $\mathbb{N}$:
\\\>\>\> ( $\forall$ i $\in$ $\mathbb{N}_k$: monitors(i, rte, c, p, p', m, n) $\wedge$ equals(s', n(k)) ) $\wedge$
\\\>\>\>[\=( eqMode(n(k), ``normal'') $\vee$ eqMode(n(k), ``compromised'')] $\wedge$
\\\>\>\>IF \=eqMode(n(k), ``normal'') THEN 
\\\>\>\>\>eqFlag(n(k), ``completed'') $\wedge$ equals(s', t') 
\\\>\>\>ELSE $\neg$ equals(s', t')
\end{tabbing}
\enlargethispage*{1cm}
In detail, the predicate says that if we execute specification ($\omega$) in an arbitrary safe pre-state (s) and environment (d) and execute program ($\kappa$) in an arbitrary pre-state (t s.t. s equals t) and environment (e$_r$ s.t. e$_r$ equals d) then there is a finite natural number (k) at which monitor can be observed such that for all iterations until k, the monitor continuous operation. However, at iteration k, either the monitor is in a "normal" mode or in a "compromised" mode. If the mode is "normal", then the component under monitoring has finished its job safely and the post-state of the program execution (t') is equal to post-state (t) of the specification execution, otherwise component is compromised and thus the program execution state (s') and specification execution state (t') are inconsistent. 
%Given the assumption that the specification always runs in a safe mode, we conclude that the program execution has been compromised.
%\enlargethispage*{1cm}
The core semantics of \emph{monitor} is captured by an auxiliary predicate \emph{monitors} that is defined as a relation on
\begin{itemize}
\item the number of observation $i$ w.r.t. of a component,
\item an observation (run-time event) $rte$, component $c$ being observed,
%\item component $c$ being observed,
\item sets of pre- and post-environments $e$ and $e'$ resp. and
%\item a set of post-environments $e'$,
\item sets of pre- and post-states $s$ and $s'$ respectively.
%\item a set of post-states $s'$.
\end{itemize}

The predicate \emph{monitors} is formalized as follows:
\begin{tabbing}
\textbf{monitors}  $\subseteq$ $\mathbb{N}$ \=$\times$ \textbf{RTEvent} $\times$ \textbf{Component} \\\>$\times$ \textbf{Environment}$^*$ $\times$ \textbf{Environment}$^*$
\\\>$\times$ \textbf{State}$^*$ $\times$ \textbf{State}$_\bot^*$
\\monitors(i, \textlbrackdbl rte\textrbrackdbl, \textlbrackdbl c\textrbrackdbl, e, e', s, s') $\Leftrightarrow$
\\ eqMode(s(i), "completed")
\\ $\vee$
\\ {[} ( \=eqMode(s(i), ``running'') $\vee$  eqMode(s(i), ``ready'') )  $\wedge$ 
\\\>$\neg$ eqMode(s(i), "compromised") $\wedge$ \textlbrackdbl c\textrbrackdbl(e(i))(e'(i), s(i), s'(i)) $\wedge$
\\\> $\exists$ \=oe $\in$ ObEvent: equals(rte, store(\textlbrackdbl name(rte)\textrbrackdbl)(e(i))) $\wedge$ 
\\\>\> {\tt run}({\tt type}(oe, c), c, e, e', s, s', i, eqMode(s', "normal"))) {]}
%\\\> IF \= entryEvent(oe, c) THEN
%\\\>\> data(c, s(i), s'(i)) $\wedge$ 
%\\\>\> [ [ \=preconditions(c, e(i), e'(i), s(i), s'(i), ``compromised'') $\wedge$  
%\\\>\>\>equals(s(i+1), s(i)) $\wedge$ equals(s'(i+1), s(i+1)) $\wedge$ 
%\\\>\>\> setFlag(inState(s'(i+1)), ``compromised'') ] $\vee$
%\\\>\> [ \=preconditions(c, e(i), e'(i), s(i), s'(i), ``normal'') $\wedge$ 
%\\\>\>\> setMode(s(i), ``running'') $\wedge$ 
%\\\>\>\> LET \= cseq = components(c) IN
%\\\>\>\>\> equals(s(i+1), s'(i)) $\wedge$ equals(e(i+1), e'(i)) $\wedge$
%\\\>\>\>\>$\forall$ \= c$_1$ $\in$ cseq, rte$_1$ $\in$ RTEvent: 
%\\\>\>\>\>\> arrives(rte$_1$, s(i+1)) $\wedge$ 
%\\\>\>\>\>\> monitor(i+1, rte$_1$, c$_1$, e(i+1), e'(i+1), s(i+1), s'(i+1))
%\\\>\> END ] ]
%\\\> ELSE IF \= exitEvent(oe, c) THEN
%\\\>\> data(c, s(i), s'(i)) $\wedge$ eqMode(inState(s'(i)), ``completed'') $\wedge$ 
%\\\>\> [ [ \=postconditions(c, e(i), e'(i), s(i), s'(i), ``compromised'') $\wedge$  
%\\\>\>\> equals(s(i+1), s(i)) $\wedge$ equals(s'(i+1), s(i+1)) $\wedge$ 
%\\\>\>\> setFlag(inState(s'(i+1)), ``compromised'') ] $\vee$
%\\\>\> [ \=postconditions(c, e(i), e'(i), s(i), s'(i), ``normal'') $\wedge$
%\\\>\>\> equals(s(i+1), s'(i)) $\wedge$ equals(e(i+1), e'(i)) $\wedge$ 
%\\\>\>\> setMode(inState(s'(i+1), ``completed'') ] ]
%\\\> ELSE IF \= allowableEvent(oe, c) THEN
%\\\>\> equals(s(i+1), s'(i)) $\wedge$ equals(e(i+1), e'(i))
%\\\> ELSE \= 
%\\\>\>equals(s(i+1), s(i)) $\wedge$ equals(s'(i+1), s(i+1)) $\wedge$ 
%\\\>\> setFlag(inState(s'(i+1)), ``compromised'')
%\\\> END {]}
\end{tabbing}
In detail, the predicate \emph{monitors} is defined such that, at any arbitrary observation either the execution is completed and returns or the current execution state $s(i)$ of component $c$ is ``ready'' or ``running'' and the current execution state is safe and behavior of the component $c$ has been evaluated and there is a run-time event $oe$ that we want to observe (and thus equals an \emph{observation} $rte$) and then any of the following
can happen:
\begin{itemize}
\item either the \emph{prediction} resp. \emph{observation} is an entry event of the component $c$, then it waits until the complete data for $c$ arrives, if so, then
\begin{itemize}
\item either the preconditions and the invariant of ``normal'' behavior of the component hold and there is no potential attack for the component (as modeled by semantic rule {\tt comp} in Figure~\ref{fig:semrules}); if so, then the subnetwork of the component is initiated and the sub-components in the subnetwork are monitored iteratively with the corresponding arrival of the \emph{observation}
\item or the preconditions and the invariant of ``compromised''  behavior of the component hold or some attack plan is detected for the component, in this case the state is marked to ``compromised'' and returns
\end{itemize}
\item or the \emph{observation} is an exit event and after the arrival of complete data, the post-conditions and the invariant hold and if there is no potential threat detected, then the resulting state is marked as ``completed''
\item or the \emph{observation} is an allowable event, the invariant holds and there is no threat for $c$, then the $c$ continues the execution
\item or the \emph{observation} is an unexpected event (i.e. none of the above holds), then the state is marked as ``compromised'' and returns.
\end{itemize}
All of the above choices are modeled by the corresponding semantic inference rule of \texttt{run}, see Figure~\ref{fig:semrules}.
\subsection{Overview of the Soundness}\label{subsec:soundness}
The intent of soundness statement is to articulate whether the system's behavior is consistent with the behavioral specification. Essentially, the goal here is to show the absence of false negative alarm such that whenever the security monitor alarms there is indeed a semantic inconsistency between post-state of the program execution and post-state of the specification execution. The soundness theorem is stated as follows:

\begin{thm}[\textbf{Soundness of security monitor}]\label{soundness} 
The result of the security monitor is sound for any execution of the target system and its specification, iff, the specification is consistent\footnote[4]{See definition of the corresponding predicate \emph{consistent} in \cref{subsec:afp}.} with the program and the program executes in a safe pre-state and in an environment that is consistent with the environment of the specification, then \begin{itemize}
\item for the pre-state of the program, there is an equivalent safe pre-state for which the specification can be executed and the monitor can be observed and
\item if we execute the specification in an equivalent safe pre-state and observe the monitor at any arbitrary (combined) post-state, then 
\begin{itemize}
\item either there is no alarm, and then the post-state is safe and the program execution (post-state) is semantically consistent with the specification execution (post-state)
\item or there is an alarm, and then the post-state is compromised and the program execution (post-state) and the specification execution (post-state) are semantically inconsistent.
\end{itemize} 
\end{itemize}
\end{thm}

%(in a consistent setting of corresponding abstractions and environments of target system and its specification respectively) execution of the system in a safe pre-state (s) and environment (e$_r$) yields post-environment (e$_r$') and post-state (s') then 
%\begin{itemize}
%\item for some pre-, post-states (t and t' resp.) and environment (e$_s$), execution of the specification in a pre-state (t) and environment (e$_s$) yields post-environment (e$_s$') and post-state (t') where the two pre-states (s and t) are (semantically) equivalent and monitor can be observed arbitrarily such in a given combined environment (e$_r$;e$_s$) and pre-state (s;t) execution of the monitor yields combined post-state (s';t') and post-environment (e$_r$';e$_s$') and
%\item if execution of the specification yields post-environment (e$_s$') and post-state (t') and pre-states (s and t) are equal and the monitor is observed in a combined post-state (s';t'), then 
%\begin{itemize}
%\item either monitor does not alarm (b is \texttt{True}) then the post-states (s' and t') are (semantically) consistent 
%\item or monitor alarms (b is \texttt{False}) and then the post-states (s' and t') are semantically inconsistent (which implies that execution of the system is compromised with the assumption that the specification is safe).
%\end{itemize} 
%\end{itemize} 
%\end{thm}
Formally, soundness theorem has the following signatures and definition.

\begin{tabbing}
Soundness\_ad $\subseteq$ $\mathbb{P}$(AppImpl $\times$ AppSpec $\times$ Bool)
\\Soundness\_ad($\kappa$, $\omega$, b) $\Leftrightarrow$
\\$\forall$ \=e$_s$ $\in$ Environment$_s$, e$_r$, e$_r$' $\in$ Environment$_r$, s, s' $\in$ State$_r$:
\\\>consistent(e$_s$, e$_r$) $\wedge$ consistent($\kappa$, $\omega$) $\wedge$ 
\\\>\textlbrackdbl $\kappa$\textrbrackdbl(e$_r$)(e$_r$', s, s') $\wedge$ eqMode(s, "normal")
\\$\Rightarrow$
\\\>$\exists$ \=t, t' $\in$ State$_s$, e$_s$' $\in$ Environment$_s$: 
\\\>\>equals(s, t) $\wedge$ \textlbrackdbl $\omega$\textrbrackdbl(e$_s$)(e$_s$', t, t') $\wedge$ monitor($\kappa$, $\omega$)(e$_r$;e$_s$)(s;t, s';t')
$\wedge$
\\\>$\forall$ \=t, t' $\in$ State$_s$, e$_s$' $\in$ Environment$_s$: 
\\\>\>equals(s, t) $\wedge$ \textlbrackdbl $\omega$\textrbrackdbl(e$_s$)(e$_s$', t, t') $\wedge$ monitor($\kappa$, $\omega$)(e$_r$;e$_s$)(s;t, s';t')
\\\>$\Rightarrow$ \=
\\\> LET \=b = eqMode(s', "normal") IN
\\\>\>IF b = \texttt{True} THEN equals(s', t') ELSE $\neg$ equals(s', t')  \hspace*{0.2cm} (G)
\end{tabbing}

In detail, the soundness statement says that, if
\begin{enumerate}
\item a specification environment (e$_s$) is \emph{consistent} with a run-time environment (e$_r$) and
\item a target system ($\kappa$) is \emph{consistent} with its specification ($\omega$) and
\item in a given run-time environment (e$_r$), execution of the system ($\kappa$) transforms pre-state (s) into a post-state (s') and
\item the pre-state (s) is safe, i.e. the state is in "normal" mode,
\end{enumerate}
then 

\begin{itemize}
\item there is such pre- and post-states (t and t' respectively) and environment (e$_s$') of the specification execution such that in a given specification environment (e$_s$), execution of the specification ($\omega$) transforms pre-state (t) into a post-state (t') and
\item the pre-states s and t are \emph{equal} and \emph{monitor}ing of the system ($\kappa$) transforms combined pre-state (s;t) into a combined post-state (s';t') and if
\item in a given specification environment (e$_s$), execution of the specification ($\omega$) transforms pre-state (t) into a post-state (t') and
\item the pre-states s and t are \emph{equal} and \emph{monitor}ing of the system ($\kappa$) transforms pre-state (s) into a post-state (s') then
\begin{itemize}
\item either there is no alarm (b is \texttt{True}) and then the post-state s' of program execution is safe and the resulting states s' and t' are semantically \emph{equal}
\item or the security monitor alarms (b is \texttt{False}) and then the post-state s' of program execution is compromised and the resulting states s' and t' are semantically not \emph{equal}.
\end{itemize}
\end{itemize}

In the following section we present proof of the soundness statement.
\subsection{Proof of the Soundness}
The proof is essentially a structural induction on the elements of the specification ($\omega$) of the system ($\kappa$). We have proved only interesting case $\beta$ of the specification to show that the proof works in principle. However, the proof of the remaining parts can easily be rehearsed following the similar approach. 

The proof is based on certain lemmas (see subsection~\ref{subsec:lemmas}), which are about the relations between different elements of the system and its specification (being at different levels of abstraction). These lemmas and relations can be proved based on the defined auxiliary functions and predicates (see subsection~\ref{subsec:afp}) that are based on the method suggested by Hoare~\cite{Hoare72}. 

In the following, we start proof with induction on $\eta$.

\subsubsection{Case ($\eta$)}

We can re-write (G) as 

\begin{tabbing}
Soundness\_ad($\kappa$, $\eta$, b) $\Leftrightarrow$
\\$\forall$ \=e$_s$ $\in$ Environment$_s$, e$_r$, e$_r$' $\in$ Environment$_r$, s, s' $\in$ State$_r$:
\\\>consistent(e$_s$, e$_r$) $\wedge$ consistent($\kappa$, $\eta$) $\wedge$ 
\\\>\textlbrackdbl $\kappa$\textrbrackdbl(e$_r$)(e$_r$', s, s') $\wedge$ eqMode(s, "normal")
\\$\Rightarrow$
\\\>$\exists$ \=t, t' $\in$ State$_s$, e$_s$' $\in$ Environment$_s$: 
\\\>\>equals(s, t) $\wedge$ \textlbrackdbl $\eta$\textrbrackdbl(e$_s$)(e$_s$', t, t') $\wedge$ 
\\\>\> monitor($\kappa$, $\eta$)(e$_r$;e$_s$)(s;t, s';t')
$\wedge$
\\\>$\forall$ \=t, t' $\in$ State$_s$, e$_s$' $\in$ Environment$_s$: 
\\\>\>equals(s, t) $\wedge$ \textlbrackdbl $\eta$\textrbrackdbl(e$_s$)(e$_s$', t, t') $\wedge$ 
\\\>\> monitor($\kappa$, $\eta$)(e$_r$;e$_s$)(s;t, s';t')
\\\>$\Rightarrow$ \=
\\\> LET \=b = eqMode(s', "normal") IN
\\\>\>IF b = \texttt{True} THEN equals(s', t') 
\\\>\> ELSE $\neg$ equals(s', t')   \hspace*{2cm} (G-1)
\end{tabbing}

Here, we have two syntactic cases for $\eta$ but we will show only one case in the following subsection.

\subsubsection{Case when $\eta$ = $\beta$}

\noindent We can re-write (G-1) as

\begin{tabbing}
Soundness\_ad($\kappa$, $\beta$, b) $\Leftrightarrow$
\\$\forall$ \=e$_s$ $\in$ Environment$_s$, e$_r$, e$_r$' $\in$ Environment$_r$, s, s' $\in$ State$_r$:
\\\>consistent(e$_s$, e$_r$) $\wedge$ consistent($\kappa$, $\beta$) $\wedge$ 
\\\>\textlbrackdbl $\kappa$\textrbrackdbl(e$_r$)(e$_r$', s, s') $\wedge$ eqMode(s, "normal")
\\$\Rightarrow$
\\\>$\exists$ \=t, t' $\in$ State$_s$, e$_s$' $\in$ Environment$_s$: 
\\\>\>equals(s, t) $\wedge$ \textlbrackdbl $\beta$\textrbrackdbl(e$_s$)(e$_s$', t, t') $\wedge$ 
\\\>\> monitor($\kappa$, $\beta$)(e$_r$;e$_s$)(s;t, s';t')
$\wedge$
\\\>$\forall$ \=t, t' $\in$ State$_s$, e$_s$' $\in$ Environment$_s$: 
\\\>\>equals(s, t) $\wedge$ \textlbrackdbl $\beta$\textrbrackdbl(e$_s$)(e$_s$', t, t') $\wedge$ 
\\\>\> monitor($\kappa$, $\beta$)(e$_r$;e$_s$)(s;t, s';t')
\\\>$\Rightarrow$ \=
\\\> LET \=b = eqMode(s', "normal") IN
\\\>\>IF b = \texttt{True} THEN equals(s', t') 
\\\>\> ELSE $\neg$ equals(s', t')  \hspace*{2cm} (F.1)
\end{tabbing}

\noindent From (F.1), we know

\begin{itemize}
\item consistent(e$_s$, e$_r$)   \hspace*{4cm} (1)
\item consistent($\kappa$, $\beta$)  \hspace*{4.2cm} (2)
\item \textlbrackdbl $\kappa$\textrbrackdbl(e$_r$)(e$_r$', s, s')   \hspace*{4cm} (3)
\item eqMode(s, "normal")   \hspace*{3.52cm} (4)
\end{itemize}

\noindent We show
\begin{itemize}
\item \begin{tabbing}$\exists$ \=t, t' $\in$ State$_s$, e$_s$' $\in$ Environment$_s$: equals(s, t) $\wedge$ 
\\\>\textlbrackdbl $\beta$\textrbrackdbl(e$_s$)(e$_s$', t, t') $\wedge$ monitor($\kappa$, $\beta$)(e$_r$;e$_s$)(s;t, s';t')  \hspace*{0.3cm} (G-1.1)\end{tabbing}
\item \begin{tabbing}$\forall$ \=t, t' $\in$ State$_s$, e$_s$' $\in$ Environment$_s$: equals(s, t) $\wedge$ 
\\\>\textlbrackdbl $\beta$\textrbrackdbl(e$_s$)(e$_s$', t, t') $\wedge$ monitor($\kappa$, $\beta$)(e$_r$;e$_s$)(s;t, s';t')
\\\>$\Rightarrow$ \=
\\\> LET \=b = eqMode(s', "normal") IN
\\\>\>IF b = \texttt{True} THEN equals(s', t') 
\\\>\> ELSE $\neg$ equals(s', t')   \hspace*{2cm} (G-1.2) \end{tabbing}
 \end{itemize}
 
 \subsubsection*{Goal: G-1.1}

\noindent We split the goal (G-1.1) into following three sub-goals:

%\begin{center}
equals(s, t)  \hspace*{4.2cm}(G-1.1.1)
%\end{center}

%\begin{center}
\textlbrackdbl $\beta$\textrbrackdbl(e$_s$)(e$_s$', t, t')  \hspace*{3.3cm}(G-1.1.2)
%\end{center}

%\begin{center}
monitor($\kappa$, $\beta$)(e$_r$;e$_s$)(e$_r$';e$_s$', s;t, s';t')  \hspace*{0.4cm}(G-1.1.3)
%\end{center}

\subsubsection*{Sub-Goal: G-1.1.1}

\noindent We define

\begin{center}
t := constructs(s, $\beta$)  \hspace*{3cm} (5)
\end{center}

\noindent We instantiate Lemma (1) with 
s as s, t as t, $\omega$ as $\beta$ to get
\begin{center}
t := constructs(s, $\beta$) $\Rightarrow$ equals(s, t) \hspace*{2cm} (I.1)
\end{center}

\noindent The goal (G-1.1.1) follows from (I.1) and definition (5). $\square$

\subsubsection*{Sub-Goal: G-1.1.2}

\noindent We expand definition (2) and get

\begin{center}
\begin{tabbing}
$\forall$ \=m, m' $\in$ State, n, n' $\in$ Environment: 
\\\>\textlbrackdbl \=$\kappa$\textrbrackdbl(n)(n', m, m') $\wedge$ eqMode(m, "normal) 
\\\>\>$\Rightarrow$ \textlbrackdbl $\beta$\textrbrackdbl(n)(n', m, m')  \hspace*{4cm} (F.2)
\end{tabbing}
\end{center}

\noindent We instantiate formula (F.2) with
m as s;t, m' as s';t', n as e$_r$;e$_s$', n' as e$_r$';$_s$' and $\kappa$ with $\kappa$ to get

\begin{center}
\begin{tabbing}
\textlbrackdbl \=$\kappa$\textrbrackdbl(e$_r$;e$_s$)(e$_r$';e$_s$', s;t, s';t') $\wedge$ eqMode(s;t, "normal") 
\\\>$\Rightarrow$ \textlbrackdbl $\beta$\textrbrackdbl(e$_r$;e$_s$)(e$_r$';e$_s$', s;t, s';t')   \hspace*{3cm} (I.2)
\end{tabbing}
\end{center}

\noindent We instantiate Lemma (4) with
s as s, s' as s', t as t, t' as t', e$_r$ as e$_r$, e$_r$' as e$_r$', e$_s$ as e$_s$, e$_s$' as e$_s$', $\kappa$ as $\kappa$
and get

\begin{center}
\begin{tabbing}
\textlbrackdbl $\kappa$\textrbrackdbl(e$_r$;e$_s$)(e$_r$';e$_s$', s;t, s';t') $\Leftrightarrow$ \textlbrackdbl $\kappa$\textrbrackdbl(e$_r$)(e$_r$', s, s')   \hspace*{0.8cm} (I.3)
\end{tabbing}
\end{center}

\noindent We instantiate Lemma (6) with
s as s, t as t, t' and get

\begin{center}
eqMode(s;t, "normal") $\Leftrightarrow$ eqMode(s, "normal")   \hspace*{0.4cm} (I.4)
\end{center}

\noindent From (I.2) with assumptions (3), (4), (I.3) and (I.4) we get

\begin{center}
\textlbrackdbl $\beta$\textrbrackdbl(e$_r$;e$_s$)(e$_r$';e$_s$', s;t, s';t')   \hspace*{3cm} (I.2')
\end{center}

\noindent We instantiate Lemma (5) with
s as s, s' as s', t as t, t' as t', e$_r$ as e$_r$, e$_r$' as e$_r$', e$_s$ as e$_s$, e$_s$' as e$_s$', $\omega$ as $\beta$
and get

\begin{center}
\begin{tabbing}
\textlbrackdbl $\beta$\textrbrackdbl(e$_r$;e$_s$)(e$_r$';e$_s$', s;t, s';t') $\Leftrightarrow$ \textlbrackdbl $\beta$\textrbrackdbl(e$_r$)(e$_r$', s, s')   \hspace*{0.8cm} (I.5)
\end{tabbing}
\end{center}

\noindent The goal (G-1.1.2) follows from (I.5) with assumption (I.2').$\square$.

\subsubsection*{Sub-Goal: G-1.1.3}

\noindent We instantiate induction assumption (on $\eta$) with
$\kappa$ as $\kappa$, $\omega$ as $\beta$, b as b to get
\begin{tabbing}
\\$\forall$ \=e$_s$ $\in$ Environment$_s$, e$_r$, e$_r$' $\in$ Environment$_r$, s, s' $\in$ State$_r$:
\\\>consistent(e$_s$, e$_r$) $\wedge$ consistent($\kappa$, $\beta$) $\wedge$ 
\\\>\textlbrackdbl $\kappa$\textrbrackdbl(e$_r$)(e$_r$', s, s') $\wedge$ eqMode(s, "normal")
\\$\Rightarrow$
\\\>$\exists$ \=t, t' $\in$ State$_s$, e$_s$' $\in$ Environment$_s$: 
\\\>\>equals(s, t) $\wedge$ \textlbrackdbl $\beta$\textrbrackdbl(e$_s$)(e$_s$', t, t') $\wedge$ 
\\\>\>monitor($\kappa$, $\beta$)(e$_r$;e$_s$)(s;t, s';t')
$\wedge$
\\\>$\forall$ \=t, t' $\in$ State$_s$, e$_s$' $\in$ Environment$_s$: 
\\\>\>equals(s, t) $\wedge$ \textlbrackdbl $\beta$\textrbrackdbl(e$_s$)(e$_s$', t, t') $\wedge$ 
\\\>\>monitor($\kappa$, $\beta$)(e$_r$;e$_s$)(s;t, s';t')
\\\>$\Rightarrow$ \=
\\\> LET \=b = eqMode(s', "normal") IN
\\\>\>IF b = \texttt{True} THEN equals(s', t') 
\\\>\> ELSE $\neg$ equals(s', t')   \hspace*{2cm} (I.6)
\end{tabbing}

\noindent We instantiate (I.6) with
e$_s$ as e$_s$, e$_s$' as e$_s$', e$_r$ as e$_r$, e$_r$' as e$_r$', s as s, s' as s' to get

\begin{tabbing}
con\=sistent(e$_s$, e$_r$) $\wedge$ consistent($\kappa$, $\beta$) $\wedge$
\\\> \textlbrackdbl $\kappa$\textrbrackdbl(e$_r$)(e$_r$', s, s') $\wedge$ eqMode(s, "normal")
\\$\Rightarrow$
\\\>$\exists$ \=t, t' $\in$ State$_s$, e$_s$' $\in$ Environment$_s$: 
\\\>\>equals(s, t) $\wedge$ \textlbrackdbl $\beta$\textrbrackdbl(e$_s$)(e$_s$', t, t') $\wedge$ 
\\\>\>monitor($\kappa$, $\beta$)(e$_r$;e$_s$)(s;t, s';t')
$\wedge$
\\\>$\forall$ \=t, t' $\in$ State$_s$, e$_s$' $\in$ Environment$_s$: 
\\\>\>equals(s, t) $\wedge$ \textlbrackdbl $\beta$\textrbrackdbl(e$_s$)(e$_s$', t, t') $\wedge$ 
\\\>\>monitor($\kappa$, $\beta$)(e$_r$;e$_s$)(s;t, s';t')
\\\>$\Rightarrow$ \=
\\\> LET \=b = eqMode(s', "normal") IN
\\\>\>IF b = \texttt{True} THEN equals(s', t') 
\\\>\>ELSE $\neg$ equals(s', t')  \hspace*{2cm} (I.6.1)
\end{tabbing}

\noindent The goal (G-1.1.3) follows from (I.6.1) with assumptions (1), (2), (3), (4). Hence goal (G-1.1) is proved. $\square$

\subsubsection*{Goal: G-1.2}

\noindent We know

\begin{itemize}
\item equals(s, t)  \hspace*{3cm} (6)
\item \textlbrackdbl $\beta$\textrbrackdbl(e$_s$)(e$_s$', t, t')  \hspace*{2.2cm} (7)
\item monitor($\kappa$, $\beta$)(e$_r$)(e$_r$', s, s')  \hspace*{0.9cm} (8)
\end{itemize}

\noindent We show
\begin{center}
\begin{tabbing}
LET \=b = eqMode(s', "normal") IN
\\\>IF b = \texttt{True} THEN equals(s', t') 
\\\>ELSE $\neg$ equals(s', t')
 \hspace*{2cm} (G-1.2')
 \end{tabbing}
\end{center}

\noindent We have two cases here

\subsubsection*{\underbar{Case 1:  b = \texttt{True}}} 

\noindent We know

\begin{center}
eqMode(s', "normal")  \hspace*{1cm} (10)
\end{center}
\noindent We show
\begin{center}
equals(s', t')  \hspace*{2cm} (G-1.2'')
\end{center}

\noindent We define

\begin{center}
t' := constructs(s', $\beta$)  \hspace*{0.9cm} (11)
\end{center}

\noindent We instantiate Lemma (1) with
s as s', t as t' to get
\begin{center}
t' := constructs(s', $\beta$) $\Rightarrow$ equals(s', t')  \hspace*{1cm} (I.7)
\end{center}

\noindent The goal (G-1.2'') follows from (I.7) with def. (11) and (10).$\square$

\subsubsection*{\underbar{Case 2: b = \texttt{False}}}

\noindent We know

\begin{center}
$\neg$ eqMode(s', "normal")  \hspace*{1cm} (12)
\end{center}

\noindent We instantiate Lemma (7) with
s as s' and get

\begin{center}
$\neg$ eqMode(s', "normal") $\Rightarrow$ eqMode(s', "compromised")  \hspace*{0.3cm} (I.8)
\end{center}

\noindent From (I.8) with assumption (12), we know

\begin{center}
eqMode(s', "compromised")  \hspace*{0.7cm} (13)
\end{center}

\noindent We show
\begin{center}
$\neg$ equals(s', t')  \hspace*{3cm} (G-1.2''')
\end{center}

\noindent We instantiate Lemma (2) with
s as s, s' as s', t as t, t' as t', e$_r$ as e$_r$, e$_r$' as e$_r$', e$_s$ as e$_s$, e$_s$' as e$_s$', $\kappa$ as $\kappa$, $\omega$ as $\beta$ to get

\begin{tabbing}
\textlbrackdbl $\kappa$\textrbrackdbl(e$_r$)(e$_r$', s, s') $\wedge$ \textlbrackdbl $\beta$\textrbrackdbl(e$_s$)(e$_s$', t, t') 
\\$\wedge$ equals(s, t) $\wedge$ eqMode(s', ``compromised'') 
\\$\Rightarrow$ t' $\neq$ constructs(s', $\beta$)  \hspace*{3cm} (I.9)
 \end{tabbing}

\noindent From (I.9), with assumptions (3), (7), (6) and (13) we get

\begin{center}
t' $\neq$ constructs(s', $\beta$)  \hspace*{1.7cm}(14)
 \end{center}
 
\noindent We instantiate Lemma (3) with
s as s', t as t', $\omega$ as $\beta$ to get

\begin{center}
t' $\neq$ constructs(s', $\beta$) $\Rightarrow$ $\neg$ equals(s', t')  \hspace*{0.3cm} (I.10)
\end{center}

The goal (G-1.2''') follows from (I.10) with assumption (14). The proof of (G-1.2') and (G-1.2''') implies the goal (G-1.2'). $\square$

\noindent Hence, the goal (G-1.2) follows from the proofs of (G-1.2.1) and (G-1.2.2). The premise eqMode(s', "compromised") of (I.9) shows that the program execution state s' has been compromised.$\square$

%\textcolor{red}{This section also requires a little re-work.}
\subsection{Proof of the Completeness}
The proof of completeness is very similar to what we have already presented above for the soundness. However, the proof differs only for the goal (G-1.2) whose proof is presented in the previous subsection.

In the following, first we formulate the completeness theorem:

\begin{thm}[\textbf{Completeness of security monitor}]\label{completeness} 
The result of the security monitor is complete for a given execution of the target system and its specification, iff, the specification is consistent with the program and the program executes in a safe pre-state and in an environment that is consistent with the environment of the specification, then 
\begin{itemize}
\item for the pre-state of the program, there is an equivalent safe pre-state for which the specification can be executed and the monitor can be observed and
\item if we execute the specification in an equivalent safe pre-state and observe the monitor at any arbitrary (combined) post-state, then 
\begin{itemize}
\item either the program execution (post-state) is semantically consistent with the specification execution (post-state), then there is no alarm and the program execution is safe
\item or the program execution (post-state) and the specification execution (post-state) are semantically inconsistent, then there is an alarm and the program execution has been compromised.
\end{itemize} 
\end{itemize}
\end{thm}

%\begin{thm}[\textbf{Completeness of security monitor}]\label{completeness} The result of security monitor (b) is complete with respect to denotational semantics of system's specification (\textlbrackdbl $\omega$\textrbrackdbl) and target system (\textlbrackdbl $\kappa$\textrbrackdbl), iff (in a consistent setting of corresponding abstractions and environments of target system and its specification respectively) execution of the system ($\kappa$) in a safe pre-state (s) and environment (e$_r$) yields post-environment (e$_r$') and post-state (s') then 
%\begin{itemize}
%\item for some pre-, post-states (t and t' resp.) and environment (e$_s$), execution of the specification in a pre-state (t) and environment (e$_s$) yields post-environment (e$_s$') and post-state (t') where the two pre-states (s and t) are (semantically) equivalent and in a given combined environment (e$_r$;e$_s$) and pre-state (s;t) execution of the monitor yields combined post-state (s';t') and post-environment (e$_r$';e$_s$') and
%\item if execution of the specification yields post-environment (e$_s$') and post-state (t') and pre-states (s and t) are equal and the monitor yields to combined post-state (s';t'), then 
%\begin{itemize}
%\item either two post-states (s' and t') are semantically consistent and then post-state (t') is safe and there is no alarm (b is \texttt{True})
%\item or the two states are semantically inconsistent and then the post-state (t') is compromised and monitor raises flag (b is \texttt{False}).
%\end{itemize}
%\end{itemize}
%
%\end{thm}

Formally, completeness theorem has the following signatures and definition.

\begin{tabbing}
Completeness\_ad $\subseteq$ $\mathbb{P}$(AppImpl $\times$ AppSpec $\times$ Bool)
\\Completeness\_ad($\kappa$, $\omega$, b) $\Leftrightarrow$
\\$\forall$ \=e$_s$ $\in$ Environment$_s$, e$_r$, e$_r$' $\in$ Environment$_r$, s, s' $\in$ State$_r$:
\\\>consistent(e$_s$, e$_r$) $\wedge$ consistent($\kappa$, $\omega$) $\wedge$ 
\\\>\textlbrackdbl $\kappa$\textrbrackdbl(e$_r$)(e$_r$', s, s') $\wedge$ eqMode(s, "normal")
\\$\Rightarrow$
\\\>$\exists$ \=t, t' $\in$ State$_s$, e$_s$' $\in$ Environment$_s$: 
\\\>\>equals(s, t) $\wedge$ \textlbrackdbl $\omega$\textrbrackdbl(e$_s$)(e$_s$', t, t') $\wedge$ 
\\\>\>monitor($\kappa$, $\omega$)(e$_r$;e$_s$)(s;t, s';t')
$\wedge$
\\\>$\forall$ \=t, t' $\in$ State$_s$, e$_s$' $\in$ Environment$_s$: 
\\\>\>equals(s, t) $\wedge$ \textlbrackdbl $\omega$\textrbrackdbl(e$_s$)(e$_s$', t, t') $\wedge$ 
\\\>\>monitor($\kappa$, $\omega$)(e$_r$;e$_s$)(s;t, s';t')
\\\>$\Rightarrow$ \=
\\\>\>IF \=equals(s', t')  THEN 
\\\>\>\> b = \texttt{True} $\wedge$ b = eqMode(s', ``normal'')
\\\>\>ELSE b = \texttt{False} $\wedge$ b = eqMode(s', "normal")   \hspace*{0.5cm} (G')
\end{tabbing}

In detail, the completeness statement says that, if
\begin{enumerate}
\item a specification environment (e$_s$) is \emph{consistent} with a run-time environment (e$_r$) and
\item a target system ($\kappa$) is \emph{consistent} with its specification ($\omega$) and
\item in a given run-time environment (e$_r$), execution of the system ($\kappa$) transforms pre-state (s) into a post-state (s') and
\item the pre-state (s) is safe, i.e. the state is in "normal" mode,
\end{enumerate}
then 
\begin{itemize}
\item there is such pre- and post-states (t and t' respectively) and environment (e$_s$') of specification execution such that in a given specification environment (e$_s$), execution of the specification ($\omega$) transforms pre-state (t) into a post-state (t') and
\item the pre-states s and t are \emph{equal} and \emph{monitor}ing of the system ($\kappa$) transforms combined pre-state (s;t) into a combined post-state (s';t') and if
\item in a given specification environment (e$_s$), execution of the specification ($\omega$) transforms pre-state (t) into a post-state (t') and
\item the pre-states s and t are \emph{equal} and \emph{monitor}ing of the system ($\kappa$) transforms pre-state (s) into a post-state (s'), then
\begin{itemize}
\item either the resulting two post-states s' and t' are semantically \emph{equal} and there is no alarm
\item or the resulting two post-states s' and t' are semantically not \emph{equal} and then the security monitor alarms.
\end{itemize}
\end{itemize}

In the following, we discuss proof of the completeness statement.

\subsubsection{Case when $\eta$ = $\beta$}

We can re-write (G') as

\begin{tabbing}
Soundness\_ad($\kappa$, $\beta$, b) $\Leftrightarrow$
\\$\forall$ \=e$_s$ $\in$ Environment$_s$, e$_r$, e$_r$' $\in$ Environment$_r$, s, s' $\in$ State$_r$:
\\\>consistent(e$_s$, e$_r$) $\wedge$ consistent($\kappa$, $\beta$) $\wedge$ 
\\\>\textlbrackdbl $\kappa$\textrbrackdbl(e$_r$)(e$_r$', s, s') $\wedge$ eqMode(s, "normal")
\\$\Rightarrow$
\\\>$\exists$ \=t, t' $\in$ State$_s$, e$_s$' $\in$ Environment$_s$: 
\\\>\>equals(s, t) $\wedge$ \textlbrackdbl $\beta$\textrbrackdbl(e$_s$)(e$_s$', t, t') $\wedge$ monitor($\kappa$, $\beta$)(e$_r$;e$_s$)(s;t, s';t')
$\wedge$
\\\>$\forall$ \=t, t' $\in$ State$_s$, e$_s$' $\in$ Environment$_s$: 
\\\>\>equals(s, t) $\wedge$ \textlbrackdbl $\beta$\textrbrackdbl(e$_s$)(e$_s$', t, t') $\wedge$ monitor($\kappa$, $\beta$)(e$_r$;e$_s$)(s;t, s';t')
\\\>$\Rightarrow$ \=
\\\>\>IF \=equals(s', t')  THEN b = \texttt{True} $\wedge$ b = eqMode(s', ``normal'')
\\\>\>ELSE b = \texttt{False} $\wedge$ b = eqMode(s', "normal")  \hspace*{1cm} (F'.1)
\end{tabbing}

\noindent From (F'.1), we know 

\begin{itemize}
\item consistent(e$_s$, e$_r$)  \hspace*{1.5cm}(1')
\item consistent($\kappa$, $\beta$)  \hspace*{1.6cm} (2')
\item \textlbrackdbl $\kappa$\textrbrackdbl(e$_r$)(e$_r$', s, s')   \hspace*{1.4cm} (3')
\item eqMode(s, "normal")   \hspace*{0.9cm} (4')
\end{itemize}

\noindent We show
\begin{itemize}
\item \begin{tabbing}$\exists$ \=t, t' $\in$ State$_s$, e$_s$' $\in$ Environment$_s$: equals(s, t) $\wedge$ 
\\\>\textlbrackdbl $\beta$\textrbrackdbl(e$_s$)(e$_s$', t, t') $\wedge$ monitor($\kappa$, $\beta$)(e$_r$;e$_s$)(s;t, s';t')  \hspace*{0.2cm} (G'-1.1)\end{tabbing}
\item \begin{tabbing}$\forall$ \=t, t' $\in$ State$_s$, e$_s$' $\in$ Environment$_s$: equals(s, t) $\wedge$ 
\\\>\textlbrackdbl $\beta$\textrbrackdbl(e$_s$)(e$_s$', t, t') $\wedge$ monitor($\kappa$, $\beta$)(e$_r$;e$_s$)(s;t, s';t')
\\\>$\Rightarrow$ \=
\\\>\>IF \=equals(s', t')  THEN b = \texttt{True}$\wedge$b=eqMode(s', ``normal'')
\\\>\>ELSE b = \texttt{False} $\wedge$ b = eqMode(s', "normal")   \hspace*{0.3cm} (G'-1.2) \end{tabbing}
 \end{itemize}
 
\subsubsection*{Goal: G'-1.1}

The proof is similar to as for the soundness goal (G.1.1) as discussed in the subsection. $\square$

\subsubsection*{Goal: G'-1.2}

\noindent We know

\begin{itemize}
\item equals(s, t)   \hspace*{3.5cm} (5')
\item \textlbrackdbl $\beta$\textrbrackdbl(e$_s$)(e$_s$', t, t')  \hspace*{2.7cm} (6')
\item monitor($\kappa$, $\beta$)(e$_r$)(e$_r$', s, s')  \hspace*{1.4cm} (7')
\end{itemize}

\noindent We show
\begin{center}
\begin{tabbing}
IF \=equals(s', t')  THEN b = \texttt{True} $\wedge$ b = eqMode(s', ``normal'')
\\ELSE b = \texttt{False} $\wedge$ b = eqMode(s', "normal")
  \hspace*{1cm} (G'-1.2')
 \end{tabbing}
\end{center}

\noindent We have two cases here

\subsubsection*{\underbar{Case 1: equals(s', t') holds}}

\noindent We know

\begin{center}
equals(s', t')  \hspace*{1.5cm} (8')
\end{center}

\noindent We show
\begin{center}
 b = \texttt{True} $\wedge$ b = eqMode(s', "normal")  \hspace*{1cm} (G'-1.2'')
\end{center}

\noindent To prove the goal, it suffices to show

\begin{center}
eqMode(s', "normal) = \texttt{True}  \hspace*{2.5cm} (G'-1.2".1)
\end{center}

\noindent We instantiate Lemma (8) with
s as s' and t as t' and get

\begin{center}
equals(s', t') $\Rightarrow$ eqMode(s', "normal")  \hspace*{1cm} (I'.1)
\end{center}

The goal (G'.1.2".1) follows from (I'.1) with assumption (8'). Hence the goal (G'.1.2") is proved. $\square$

Furthermore, the goal shows that there is no alarm when the two post-states (s' and t') are
equivalent and are not compromised.

\subsubsection*{\underbar{Case 2: $\neg$ equals(s', t') holds}} 

\noindent We know

\begin{center}
$\neg$ eqMode(s', "normal")  \hspace*{2cm} (9')
\end{center}

\noindent We show

\begin{center}
 b = \texttt{False} $\wedge$ b = eqMode(s', "normal")  \hspace*{1cm} (G'-1.2'')
\end{center}

\noindent To prove the goal, it suffices to show

\begin{center}
eqMode(s', "normal) = \texttt{False}  \hspace*{2.3cm} (G'-1.2".1)
\end{center}

\noindent We instantiate Lemma (9) with
s as s' and t as t' and get

\begin{center}
$\neg$ equals(s', t') $\Rightarrow$ $\neg$ eqMode(s', "normal")  \hspace*{1cm} (I'.3)
\end{center}

The goal (G'.1.2".1) follows from (I'.3) with assumption (9'). Hence the goal (G'.1.2") is proved. $\square$

\noindent Furthermore, we instantiate Lemma (7) with
s as s' to get

\begin{center}
$\neg$ eqMode(s', "normal") $\Rightarrow$ eqMode(s', "compromised")  \hspace*{0cm} (I'.4)
\end{center}

\noindent From (I'.4) with the proved goal (G'.1.2") we get

\begin{center}
eqMode(s', "compromised")
\end{center}

that shows that the alarm is generated when the post-states (s' and t') are semantically not equal. Furthermore, from the assumption (2') if follows that indeed the program execution (post-state) is compromised.

%In the following subsection, we discuss auxiliary functions are predicates which are used in the proofs above.
%\textcolor{red}{This section can be an appendix. However, some of them are important for the proofs.}
\subsection{Auxiliary Functions and Predicates}\label{subsec:afp}
In this section, we declare respectively define auxiliary functions and predicates that are used in the proof of soundness and completeness above.
\begin{itemize}
\item \begin{tabbing}\textbf{constructs : State$_r$} \textbf{$\times$ AppSpec} \textbf{$\rightarrow$ State$_s$}
\\cons\=tructs(s, $\omega$) = t,
\\\> s.t. t = build($\omega$) $\wedge$ eqMode(s, "normal") $\wedge$ abstract(s, t)
\end{tabbing}
%The function ``constructs'' a specification-state (t) constructed from a given run-time state (s) and specification (mod) such that state s is safe and not compromised.
\item \begin{tabbing}\textbf{constructs : Environment$_r$} \textbf{$\times$ AppSpec} \textbf{$\rightarrow$ Environment$_s$}
\\cons\=tructs(e, $\omega$) = e', s.t. e' = build($\omega$) $\wedge$ abstract(e, e')
\end{tabbing}
%Analogous to aforementioned same function.
\item \begin{tabbing}\textbf{\_ ; \_ : State$_r$ $\times$ State$_s$ $\rightarrow$ State}
\\ s;t = \=state(\{$\langle$I:v$\rangle$ $\in$ store(s) : $\neg \exists$ $\langle$I:v'$\rangle$ $\in$ store(t)\} $\cup$ 
\\\>\{$\langle$I:v'$\rangle$ $\in$ store(t) : $\neg \exists$ $\langle$I:v$\rangle$ $\in$ store(s)\} $\cup$
\\\>\{$\langle$I:v''$\rangle$ : \=$\exists$ v'': $\langle$I:v$\rangle$ $\in$ store(s) $\wedge$ $\langle$I:v'$\rangle$ $\in$ store(t) $\wedge$ 
\\\>\>v'' = super(v, v')\}, flag(s))
\end{tabbing}
%The function ; combines two given states (s and t) and returns that. In principle, the combined state includes disjoint pairs of identifier and values in the two states as it is while shared identifiers are paired to corresponding super value between the two in the resulting state.
\item \begin{tabbing}\textbf{\_ ; \_ : Environment$_r$ $\times$ Environment$_s$ $\rightarrow$ Environment}
\\ e;e' = \=environment(\{$\langle$I:v$\rangle$ $\in$ context(e) : 
\\\>$\neg \exists$ $\langle$I:v'$\rangle$ $\in$ context(e')\} $\cup$ 
\\\>\{$\langle$I:v'$\rangle$ $\in$ context(e') : $\neg \exists$ $\langle$I:v$\rangle$ $\in$ context(e)\} $\cup$
\\\>\{$\langle$I:v''$\rangle$ :\= $\exists$ v'': $\langle$I:v$\rangle$ $\in$ context(e) $\wedge$ 
\\\>\>$\langle$I:v'$\rangle$ $\in$ context(e') $\wedge$  v'' = super(v, v')\}
\\\>\>, space(e))
\end{tabbing}
%Analogous to aforementioned same function.
\item \begin{tabbing}\textbf{super : Value$_r$ $\times$ Value$_s$ $\rightarrow$ Value}
\\super(v, v') = \= v	, if \textlbrackdbl v\textrbrackdbl $\subseteq$ \textlbrackdbl v'\textrbrackdbl
\\\> v', if \textlbrackdbl v'\textrbrackdbl $\subseteq$ \textlbrackdbl v\textrbrackdbl
\end{tabbing}
%The function "super" returns one of the two given values such that semantic of result value semantically includes other given value.
\item \begin{tabbing}\textbf{super : EnvVal$_r$ $\times$ EnvVal$_s$ $\rightarrow$ EnvVal}
\\super(v, v') = \= v	, if \textlbrackdbl v\textrbrackdbl $\subseteq$ \textlbrackdbl v'\textrbrackdbl
\\\> v', if \textlbrackdbl v'\textrbrackdbl $\subseteq$ \textlbrackdbl v\textrbrackdbl
\end{tabbing}
%Analogous to the above function.
\item \begin{tabbing}\textbf{equals $\subseteq$ $\mathbb{P}($State$_r$ $\times$ State$_s$)}
\\equals(s, t) $\Leftrightarrow$
\\ $\forall$ \=c:Component$_s$, $\omega$:AppSpec, $\kappa$: AppImpl: 
\\\>c $\in$ $\omega$ $\wedge$ c $\in$ $\kappa$ $\wedge$ \textlbrackdbl c\textrbrackdbl(e$_r$)(s, s', e$_r$') 
\\$\Rightarrow$ \=\textlbrackdbl c\textrbrackdbl(e$_s$)(t, t', e$_s$') $\wedge$ \\\>$\forall$ id: Identifier$_s$, v: Value$_s$: 
$\langle$id, v$\rangle$ $\in$ store(t) 
\\\>$\Rightarrow$
$\langle$id, v'$\rangle$ $\in$ store(s) $\wedge$ abstract(v, v')
\end{tabbing}
%The predicate``equals'' returns \texttt{True} only if for those components' execution which are also specified such that for every pair of identifier and value in latter state (t) there is 
%a corresponding pair in the former state (s) where the two values of abstract to each other. 
\item \begin{tabbing}\textbf{consistent $\subseteq$ $\mathbb{P}$(Environment$_r$ $\times$ Environment$_s$)}
\\consistent(e$_r$, e$_s$) $\Leftrightarrow$
\\ $\forall$ \=id:Identifier, v: Value$_s$, v': Value$_r$: 
\\\>$\langle$id, v$\rangle$ $\in$ context(e$_s$) $\Rightarrow$ $\langle$id, v'$\rangle$ $\in$ context(e$_r$) $\wedge$ abstract(v, v')
\end{tabbing}
%The predicate ``consistent'' returns \texttt{True} only if for every pair of identifier and environment value in latter environment there is a corresponding pair in the former environment.
\item \begin{tabbing}\textbf{consistent $\subseteq$ $\mathbb{P}$(AppImpl $\times$ AppSpec)}
\\consistent(\=$\kappa$, $\omega$) $\Leftrightarrow$ the safe execution of "$\kappa$" meets "$\omega$" and "$\omega$" 
\\always executes in a safe state, that can be formulated as follows:
\\ $\forall$ \=s, s' $\in$ State, e, e' $\in$ Environment: 
\\\>\textlbrackdbl $\kappa$\textrbrackdbl(e)(e', s, s') $\wedge$ eqMode(s, "normal") $\Rightarrow$ \textlbrackdbl $\omega$\textrbrackdbl(e)(e', s, s') $\wedge$
\\$\forall$ \=t, t' $\in$ State$_s$, d, d' $\in$ Environment$_s$: 
\\\>\textlbrackdbl $\omega$\textrbrackdbl(d)(d', t, t') $\wedge$ eqMode(t, "normal") $\Rightarrow$ eqMode(t', "normal")
\end{tabbing}
Semantically, the predicate ``consistent'' returns \texttt{True} iff only such pair of states (s and s') are related by "$\kappa$" which is also related by "$\omega$". Here the states and environment are combined of two corresponding abstractions of specification and implementation respectively. Furthermore, execution of "$\omega$" in a safe pre-state always yields a safe post-state.
\item \begin{tabbing}\textbf{abstract $\subseteq$ $\mathbb{P}$(State$_r$ $\times$ State$_s$)}
\\abstract(s, t) $\Leftrightarrow$
\\$\forall$ \=i:Identifier, v:Value$_s$: 
\\\>$\langle$i, v$\rangle$ $\in$ store(t) $\Rightarrow$ $\exists$ v':Value$_r$:\= $\langle$i, v'$\rangle$ $\in$ store(s) $\wedge$ 
\\\>\>abstract(v, v')
\end{tabbing}
%The predicate ``abstract'' returns \texttt{True} only if for every pair of identifier and value in latter state (t), there is a corresponding pair in former state (s) when both identifiers and their values are abstract to each other correspondingly.
\item \begin{tabbing}\textbf{abstract $\subseteq$ $\mathbb{P}$(Value$_r$ $\times$ Value$_s$)}
\\abstract(v, v') $\Leftrightarrow$ 
\\$\forall$ \=$\tau$, $\tau$':Type, s:State$_r$, t:State$_s$: 
\\\>equals(s, t) $\wedge$ \textlbrackdbl v\textrbrackdbl(s, $\tau$) $\wedge$ \textlbrackdbl v'\textrbrackdbl(t, $\tau$') $\Rightarrow$ \textlbrackdbl $\tau$'\textrbrackdbl $\subseteq$ \textlbrackdbl $\tau$\textrbrackdbl
\end{tabbing}
%The predicate ``abstract'' returns \texttt{True} only if two arbitrary run-time and specification states (s and t) are semantically equal and evaluation of given values in corresponding states yields respective types such that run-time type semantically contains specification type.
\item \begin{tabbing}\textbf{abstract $\subseteq$ $\mathbb{P}$(EnvVal$_r$ $\times$ EnvVal$_s$)}
\\abstract(v, v') $\Leftrightarrow$
\\$\forall$ \=$\tau$, $\tau$':Type, e:Environment$_s$, e':Environment$_r$: 
\\\>consistent(e, e') $\wedge$ \textlbrackdbl v\textrbrackdbl(e, $\tau$) $\wedge$ \textlbrackdbl v'\textrbrackdbl(e', $\tau$') $\Rightarrow$ \textlbrackdbl $\tau$'\textrbrackdbl $\subseteq$ \textlbrackdbl $\tau$\textrbrackdbl
\end{tabbing}
%This function is analogous to above function.
\end{itemize}

%\textcolor{red}{This section can also go away. I included for completeness of readability.}
\subsection{Lemmas}\label{subsec:lemmas}
In this section, we give definitions and corresponding proof hints of lemmas that were used in the proofs above.

\begin{lem} \end{lem}
\begin{tabbing}$\forall$ s $\in$ State$_r$, t $\in$ State$_s$: t = constructs(s) $\Rightarrow$ equals(s, t) \end{tabbing} %\end{lem}

%The lemma states that if specification execution state $t$ can be constructed from program execution state $s$ then the two states are semantically equal.
%\begin{proof}[\textbf{Proof Hints}]
%By expanding definition of function \emph{constructs}.
%\end{proof}
\begin{lem}  \end{lem}\begin{tabbing}$\forall$ \=s, s' $\in$ State$_r$, t, t' $\in$ State$_s$, 
\\\>$\kappa$ $\in$ AppImpl, $\omega$ $\in$ AppSpec, 
\\\>e$_r$, e$_r$' $\in$ Environment$_r$, e$_s$, e$_s$' $\in$ Environment$_s$:
\\\>\textlbrackdbl $\kappa$\textrbrackdbl(e$_r$)(e$_r$', s, s') $\wedge$ \textlbrackdbl $\omega$\textrbrackdbl(e$_s$)(e$_s$', t, t') 
\\\>$\wedge$ \=equals(s, t) $\wedge$ eqMode(s', ``compromised'') 
\\\>\>$\Rightarrow$ t' $\neq$ constructs(s')
 \end{tabbing}
% This lemma states that if execution of a program yields a ``compromised'' post-state $s'$ then post-state $t'$ (yielded by execution of specification) cannot be constructed from $s'$ with the fact that pre-states ($s$ and $t$) of both executions were semantically equivalent.
 \textbf{\em Proof Hints}
In principle, from a compromised program state, an equivalent specification safe state cannot be constructed because the program state may have inconsistent values for certain variables or new variables etc.
 \begin{lem} \end{lem}\begin{tabbing}$\forall$ s $\in$ State$_r$, t $\in$ State$_s$: t $\neq$ constructs(s) $\Rightarrow$ $\neg$ equals(s, t)\end{tabbing}
%This lemma states that if specification execution state $t$ cannot be constructed from a program execution state $s$, then the two are not semantically equivalent.
%\begin{proof}[\textbf{Proof Hints}]
%Trivial by deriving the definitions of \emph{constructs} and \emph{equals}.\end{proof}
\begin{lem} \end{lem} \begin{tabbing} $\forall$ \=s, s' $\in$ State, t, t' $\in$ State$_s$, 
\\\>e$_r$, e$_r$' $\in$ Environment$_r$, e$_s$, e$_s$' $\in$ Environment$_s$, 
\\\> $\kappa$ $\in$ AppImpl:
\\\>\textlbrackdbl $\kappa$\textrbrackdbl(e$_r$;e$_s$)(e$_r$';e$_s$', s;t, s';t') $\Leftrightarrow$ \textlbrackdbl $\kappa$\textrbrackdbl(e$_r$)(e$_r$', s, s')
\end{tabbing}
%This lemma states that if target system can be executed in arbitrary combined (run-time and specification based) states and environment, then
%the same program can also be executed in corresponding run-time states and environment only and conversely.
\textbf{\em Proof Hints}
The goal follows from the semantics of \emph{$\kappa$}.
\begin{lem} \end{lem}\begin{tabbing} $\forall$ \=s, s' $\in$ State, t, t' $\in$ State$_s$, 
\\\>e$_r$, e$_r$' $\in$ Environment$_r$, e$_s$, e$_s$' $\in$ Environment$_s$, 
\\\> $\omega$ $\in$ AppSpec: 
\\\>\textlbrackdbl $\omega$\textrbrackdbl(e$_r$;e$_s$)(e$_r$';e$_s$', s;t, s';t') $\Leftrightarrow$ \textlbrackdbl $\omega$\textrbrackdbl(e$_s$)(e$_s$', t, t')
\end{tabbing}
%This lemma states that if specification can be executed in arbitrary combined (run-time and specification based) states and environment, then
%the same specification can also be executed in corresponding specification related states and environment only and conversely.
%\begin{proof}[\textbf{Proof Hints}]
%Same as above for \emph{$\kappa$}.\end{proof}
\begin{lem} \end{lem} \begin{tabbing} $\forall$ \=s $\in$ State, t $\in$ State$_s$: \\\>eqMode(s;t, "normal") $\Leftrightarrow$ eqMode(s, "normal")
\end{tabbing}
%This lemma states that if an arbitrary combined (run-time and specification based) state is safe then the corresponding run-time state (s) is also safe and conversely.
%\begin{proof}[\textbf{Proof Hints}]
%Trivial and same as above for \emph{$\kappa$} and \emph{$\omega$}.\end{proof}
\begin{lem} \end{lem} 
\begin{tabbing}
$\forall$ \=s $\in$ State$_r$: \\\>$\neg$ eqMode(s', "normal") $\Leftrightarrow$ eqMode(s', "compromised")\end{tabbing}
%\begin{proof}[\textbf{Proof Hints}]
%The goal is a logical consequence of expansion of definition of \emph{State}.\end{proof}
\begin{lem}\end{lem} \begin{tabbing} $\forall$ s $\in$ State$_r$, t $\in$ State$_s$: equals(s, t) $\Rightarrow$ eqMode(s, "normal")\end{tabbing}
\textbf{\em Proof Hint}
The definition of \emph{equals} enables to show the goal. Also because of the fact, that two states are only equal if they can be constructed in a safe mode.
\begin{lem} \end{lem}

\begin{tabbing} $\forall$ \=s $\in$ State$_r$, t $\in$ State$_s$: \\\>$\neg$ equals(s, t) $\Rightarrow$ $\neg$ eqMode(s, "normal") \end{tabbing}
% \begin{proof}[\textbf{Proof Hints}]
%Same as of Lemma 8 above.\end{proof}

%\textcolor{red}{The following section needs revision.}
\section{Conclusion}\label{sec:conc}
%\begin{itemize}
%\item general outcomes and lessons learned
%\item applications of learning outcomes
%\end{itemize}
We have presented a sound and complete run-time security monitor for application software,
which avoids false alarms (positive or negative). The monitor implements run-time software
verification, comparing an executable application specification with the execution of its
implementation at run-time. Our main contribution, the proof of soundness and completeness,
establishes an {\em assume/guarantee}-based contract between the \emph{security monitor}
and its user, i.e. the designer of the application to be monitored. Specifically, if the user 
establishes the {\em assumptions} of the proof, then the monitor \emph{guarantees} to detect
all deviations of the execution’s behaviour relatively to the behaviour defined in the application
specification and will never produce any false alarm at run-time. 
Importantly, the proof strategy can be a fundamental building block for:
\begin{enumerate}
\item 
any proof that shows that an abstract description/specification (non-determinism) of a program 
is consistent with its concrete description/implementation (determinism/instance),
\item 
transformation rules to automatically generate sound and complete monitors (for program execution) 
from specification and
\item 
developing proof tactics to prove such tedious goals semi-automatically, significantly reducing human effort.
%\item inferring direct knowledge to make such proof easier. 
\end{enumerate}
Our future work includes the mechanisation of this proof in a proof assistant, specifically Coq, targeting
the development of a generic library based on our proof strategy so that the proof can be applied to any 
given specification and implementation.

\end{document}